\documentclass[11pt,a4paper,twoside]{paper}
\usepackage[T1]{fontenc}
\usepackage{amsmath,amssymb,amsthm}
\usepackage{graphicx}
\usepackage{parskip}
\paragraphfont{\sffamily\itshape}

\newtheoremstyle{etoile}{\parskip}{\parskip}{\itshape}
                        {0pt}{\bfseries\sffamily}{.}{ }{}
\theoremstyle{etoile}

\newtheorem{prop}{Proposition}[section]

\newtheorem{lem}[prop]{Lemma}

\makeatletter
\@addtoreset{equation}{section}
\makeatother

\newcommand\egaldef{\stackrel{\mbox{\upshape\tiny def}}{=}}

\newcommand\1{\leavevmode\hbox{\rm \small1\kern-0.35em\normalsize1}}
\newcommand\ind[1]{\1_{\{#1\}}}        
\newcommand\EE{\mathsf{E}}
\newcommand{\be}{\begin{equation}}
\newcommand{\ee}{\end{equation}}
\newcommand{\bea}{\begin{eqnarray}}
\newcommand{\eea}{\end{eqnarray}}
\def\DD{\displaystyle} 
\begin{document}

\title{Dynamical Windings of Random  Walks  and Exclusion Models.\\
 Part I: Thermodynamic Limit}

\author{Guy Fayolle \and Cyril Furtlehner }

\date{\today}

\maketitle

\abstract{We consider a system consisting of a planar random walk on
a square lattice, submitted to stochastic elementary local
deformations. Both numerical and theoretical results are reported.
Depending on the deformation transition rates, and specifically on a
parameter $\eta$ which breaks the symmetry between the left and right
orientation, the winding distribution of the walk is modified, and the
system can be in three different phases: folded, stretched and glassy.
An explicit mapping is found, leading to consider the system as a
coupling of two exclusion processes: particles of the first one move
in a landscape defined by particles of the second one, and vice-versa.
This can be viewed as an inhomogeneous exclusion process. For all
closed or periodic initial sample paths, a convenient scaling permits
to show a convergence in law (or almost surely on a modified
probability space) to a continuous curve, the equation of which is
given by a system of two non linear stochastic differential equations.
The deterministic part of this system is explicitly analyzed via
elliptic functions. In a similar way, by using a formal fluid limit
approach, the dynamics of the system is shown to be equivalent to a
system of two coupled Burgers' equations.}

\section{\bf Introduction}\label{Introduction}
Random walks are fundamental objects arising in probability. Also they
are of primary importance in various fields of physics, especially
with regard to polymers \cite{PGDG,EDWARD} and biology. For instance,
planar random walks can be used as a representation of DNA coding,
since the sequence of the four different kinds of codons (A,G) for
purines and (T,C) for pyrimidines can be considered as a random walk
on a square lattice: as a rule, (G,C) code the upward and downward
jumps, whereas (A,T) code the left and right steps \cite{BGS,CD}. It
seems therefore interesting to consider random geometrical objects as
complex systems, and to submit them to some dynamical principles, the
goal being to develop methods and tools which hopefully might be used
to tackle more realistic models.

In this context we will analyze the evolution of an arbitrary sample
path $\mathbf{C}_{N}$ of length $N$, generated by a \emph{simple}
random walk in the square lattice $\mathbf{Z}^{2}$, and subject to
local transformations. This stochastic object has a rich structure,
plays an important role in probability theory and lends itself to
sufficiently wide but non trivial results.

\begin{figure}[htb]
\begin{center}
\resizebox*{!}{5cm}{ \input{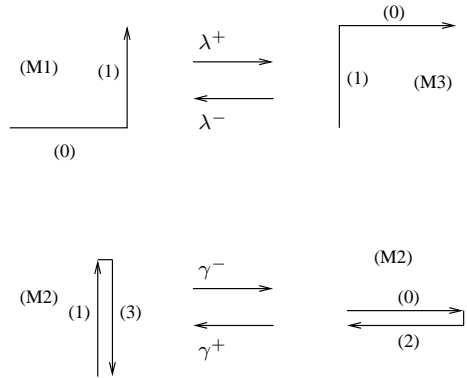}}
\caption{\label{trans} Pattern transition rates.}
\end{center}
\end{figure}

At time $t=0$, $\mathbf{C}_{N}$ is given, and we assume it has been
uniformly generated.  This means precisely that each successive jump
(up, down, left and right) building $\mathbf{C}_{N}$ is selected with
the same probability $1/4$.  Eventually $\mathbf{C}_{N} $ can be
constrained to be closed or to have fixed extremities.  Once the
initial configuration is defined, the system evolves according to the
four local pattern transformations depicted in Figure~\ref{trans}. 
Only a single point of the walk can be moved at a time, with the
constraint that no link be broken (i.e. the walk remains always
connected).

Geometrically, these patterns can be expressed as
\[
\begin{cases}
 \textrm{left bend} \ M1, \ \textrm{right bend}\ M3, \\[0.2cm]
 \textrm{vertical or horizontal fold} \ M2, \\[0.2cm]
\textrm{straight} \  (\to\to)  \ M4 , 
\end{cases}  
\]
and the following  local distortions can occur:
\[
\begin{cases} 
M1 \to M3, \ \textrm{with rate}\ \lambda^+ ,\\[0.2cm] M3 \to M1, \
\textrm{with rate}\ \lambda^- ,\\[0.2cm] 
\textrm{rotation of \emph{M2} of
angle} \pm\frac{\pi}{2} ,\ \textrm{with rate}\ \gamma^{\pm}.
\end{cases}  
\]
Hence we have defined a global Markovian continuous time evolution of
the system, with exponentially distributed jump times, the state space
of the underlying Markov chain being the set of $4^{N}$ sample paths
(or curves) $\mathbf{C}_{N}$ introduced above.

This model is somehow a kind of discrete analogue of the Rouse chain
\cite{ROUSE}, which is a popular model for polymer dynamics. There,
each point of the chain is harmonically bound to its nearest neighbor,
and move randomly in space. Some interesting statements can be made
concerning the winding properties of such chains in 2d \cite{DESBOIS}.
In this respect, from a probabilistic point of view, we keep in mind
that winding variables are for Brownian curves, and they give rise to
striking limit laws under convenient scalings (see e.g.
\cite{SPITZER,LEVY,YOR}).

The paper is organized as follows. Section $2$ presents the basic
numerical and qualitative results, which rely on \emph{ad hoc}
discrete event simulation experiments together with a convenient
graphical interface. In section $3$, we propose several possible ways
of coding the system with the related probabilistic descriptions (this
section can be only glanced at without too much damage). The last part
of the paper (section $4$) contains the main quantitative results:
relationships to exclusions models, scaling and thermodynamical limit,
fluctuation analysis via theorems of central limit type.

\section{\bf Numerical experiments}
\subsection{Observations}
\begin{figure}[htb]
\begin{center}
\includegraphics[width=0.9\textwidth]{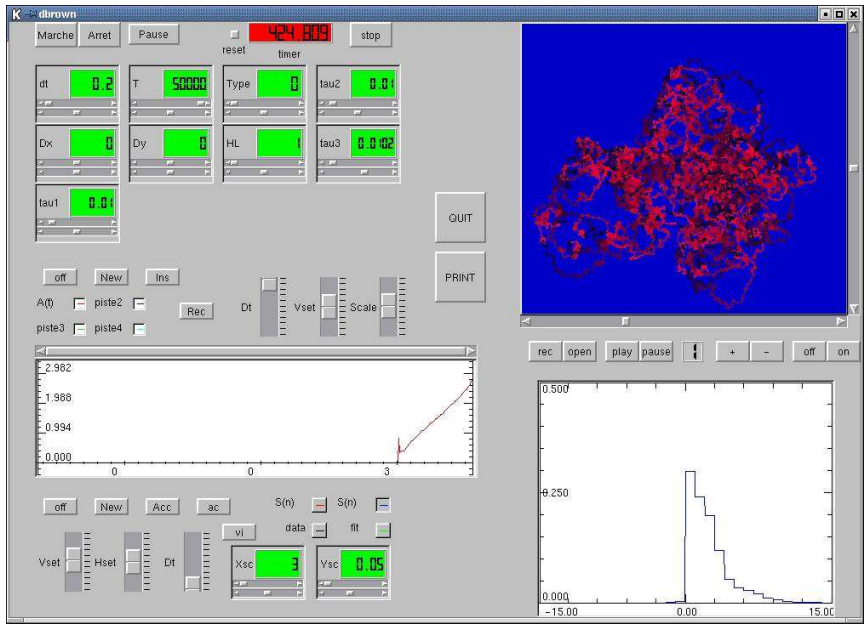}
\caption{\label{plateform}Platform for numerical experiments.}
\end{center}
\end{figure}

   We shall begin our study with several basic numerical observations.
   The model is purely stochastic and hence well-suited for
   Monte-Carlo simulations. These have been performed with the help of
   a graphical interface shown in Figure~\ref{plateform}, which
   facilitates the exploration of the relevant parameter range,
   together with the display of the main regimes and phases of the
   system.

Several parameters have to be tuned: the number of steps $N$; the
relative position of the extremities of the walk $(D_x,D_y)$; the
boundary conditions, which either can be defined to let the end points
move independently, or can be fixed or to tied by some periodic
boundary conditions; finally, the time constants associated with the
elementary transformations, $\tau_1$ with $\gamma$, $\tau_2$ and
$\tau_3$ with $\lambda^+$ and $\lambda^-$. Any walk fulfilling these
conditions is randomly generated at time $t=0$ and then evolves
stochastically, with a movement depending on the rates and boundary
conditions given above. After each new event, time is incremented by
an amount inversely proportional to the number of all possible moves
weighted by their respective rates.
\begin{figure}[htb]
\begin{center}
\begin{tabular}{cc}
\includegraphics[width=0.4\textwidth]{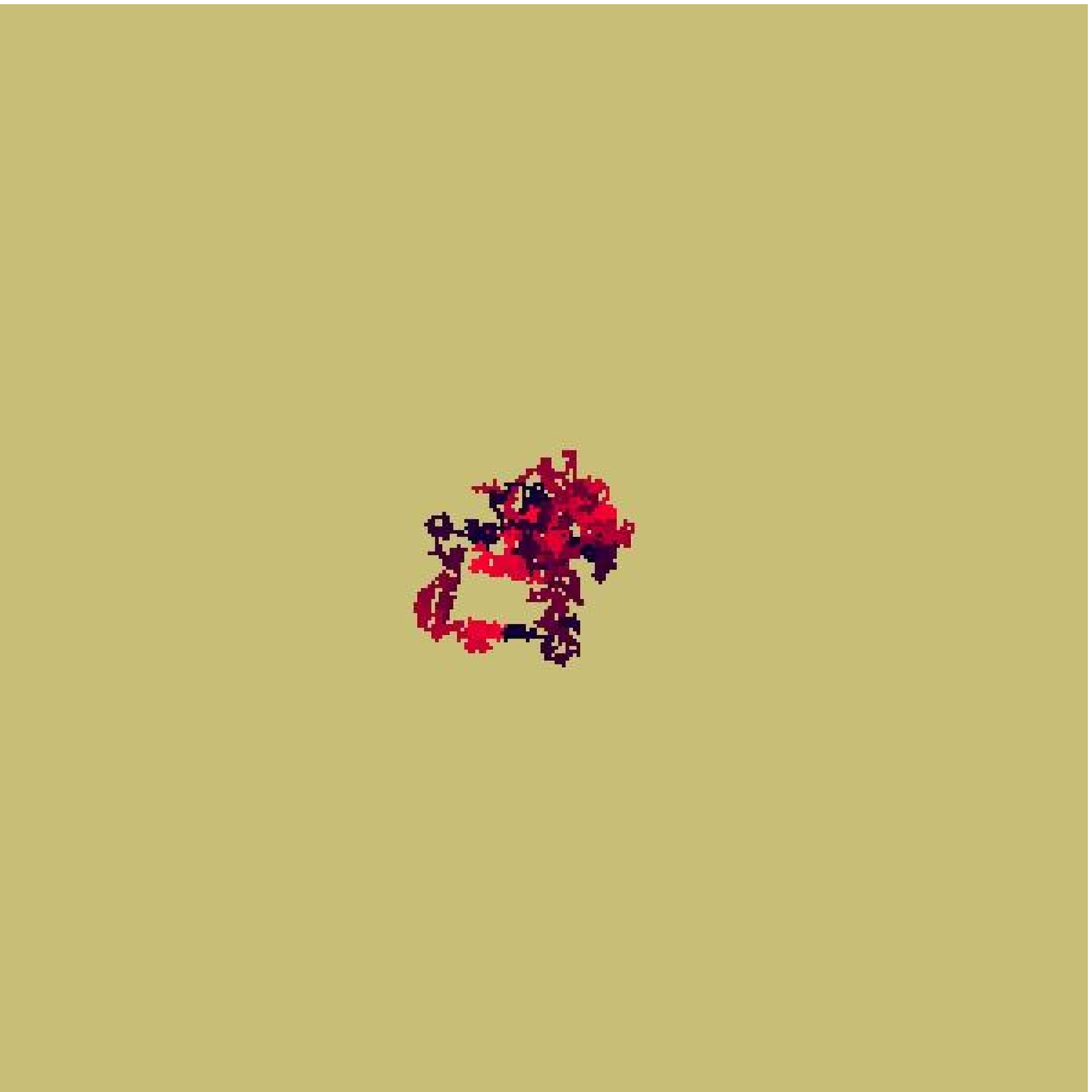}&
\includegraphics[width=0.4\textwidth]{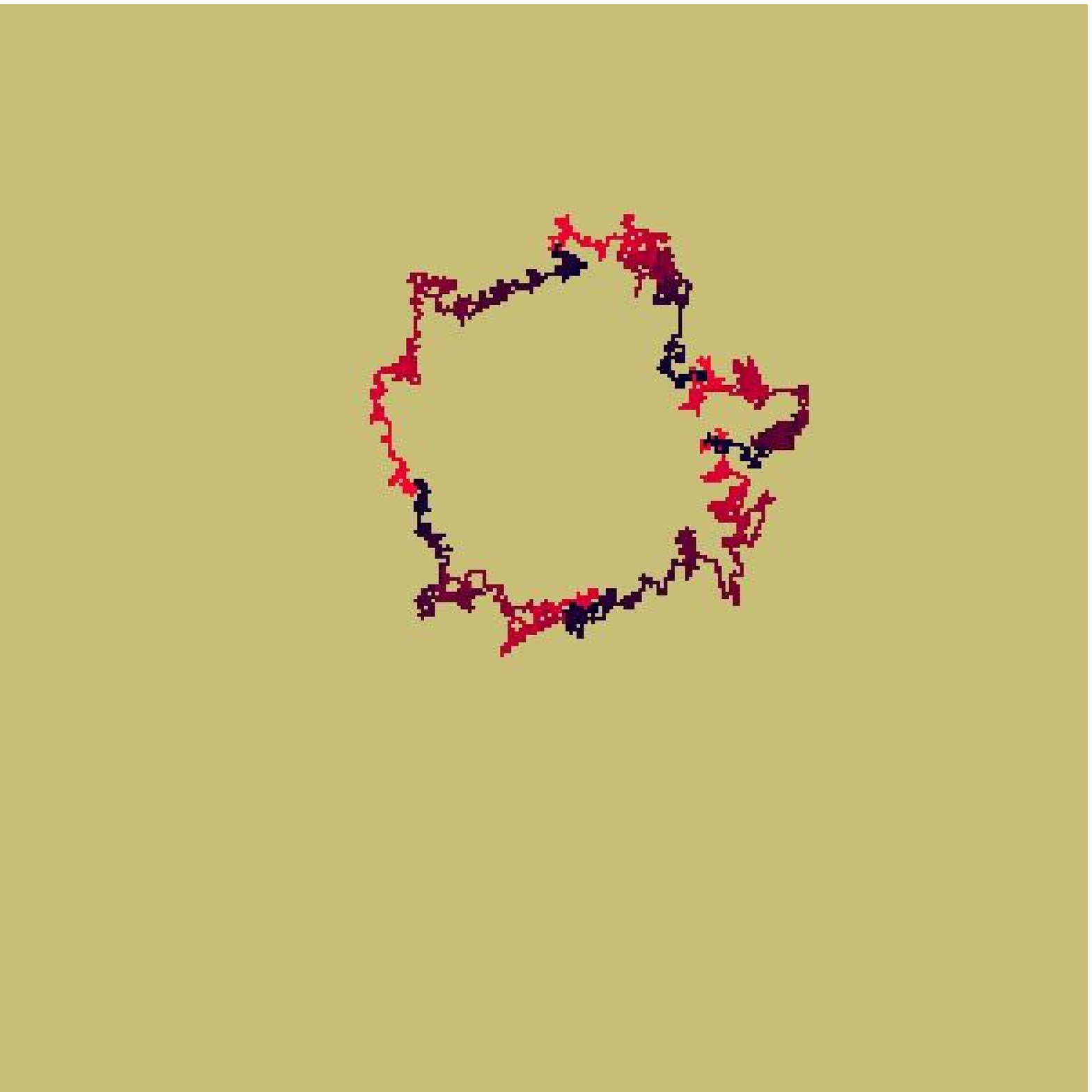}\\
(a)&(b)\\[0.2cm]
\includegraphics[width=0.4\textwidth]{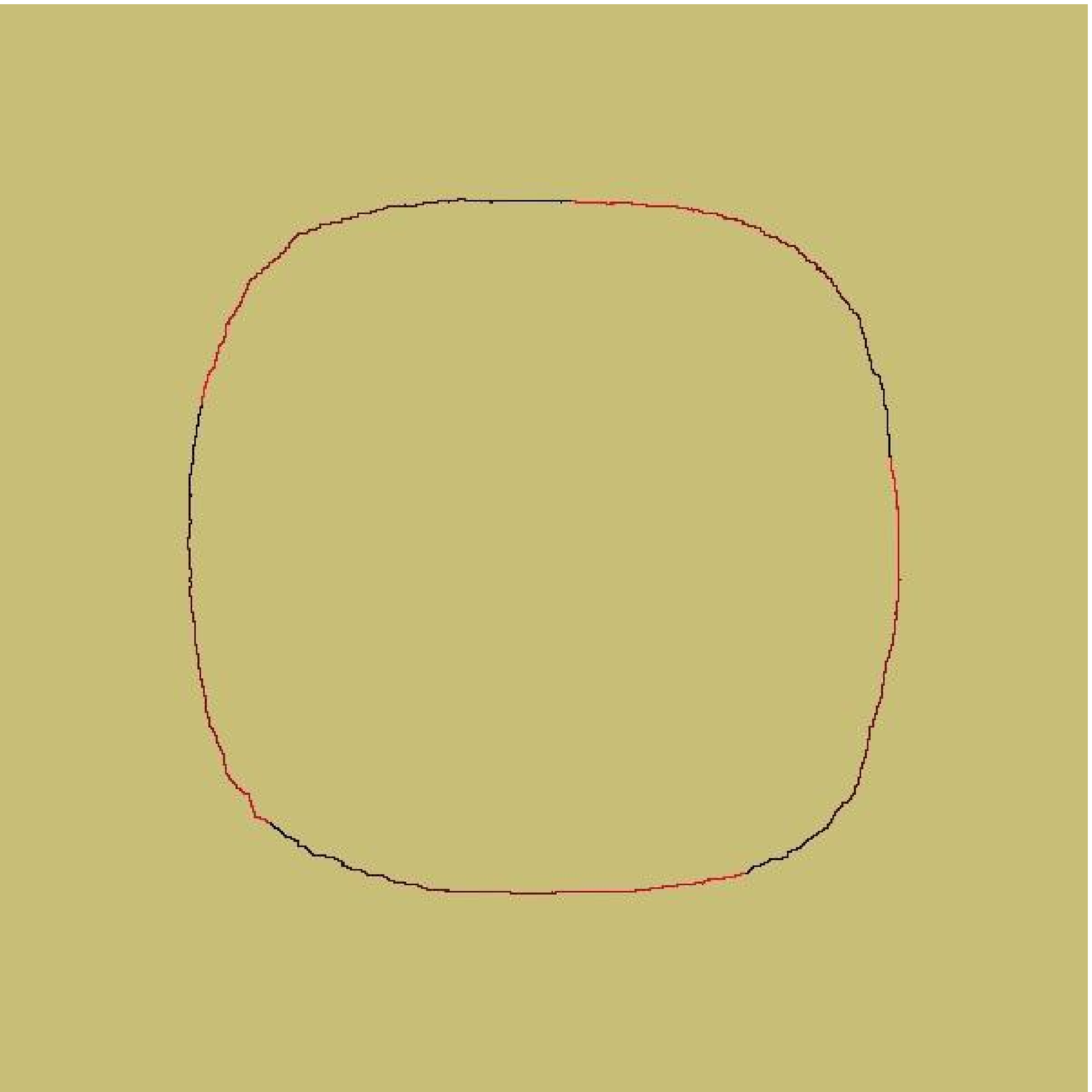}&
\includegraphics[width=0.4\textwidth]{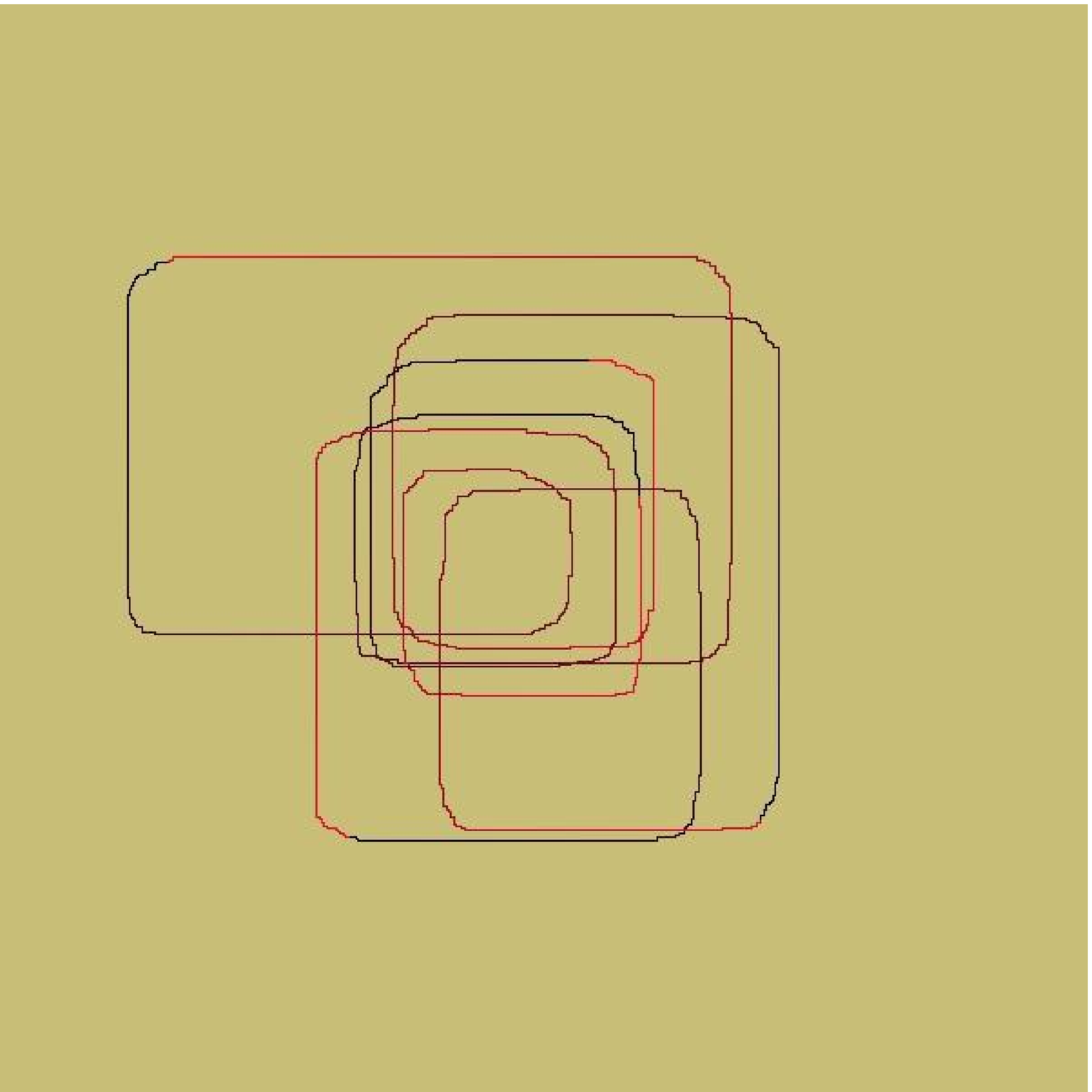}\\
(c)&(d)
\end{tabular}
\caption{\label{phases}Picture of a random walk of $N=5000$ steps,
showing the phases of the system for several values of $\eta$. Each
colored segment represents $1000$ steps. (a) $\eta=0$ (the basic
scale). (b) $\eta=5$ (scale$=$1). (c) $\eta=12.5$ (scale $= 1/6$). (d)
$\eta=250$ (scale $=1/2$).}
\end{center}
\end{figure}
Interesting things happen when we break the chiral symmetry by
imposing a \emph{detuning} between $\lambda^+$ and $\lambda^-$, in a
proportion of order
\begin{equation}\label{eta}
\eta=N\frac{\lambda^+-\lambda^-}{\lambda^++\lambda^-}.
\end{equation} 
For closed walks, four different situations can roughly be
observed (see Figure~\ref{phases}).
\begin{itemize}
\item[(a)] $\eta\lesssim 1$. The initial
configuration belongs to the equilibrium set of typical
configurations, only fluctuations are altered by the finite value of
$\eta$. 

\item[(b)] $1\lesssim\eta\lesssim6$. The system reaches an
equilibrium which still corresponds to a random walk, the fractal
dimension remaining equal to two, but a macroscopic circular drift is
observed, yielding a sort of smoking ring.

\item[(c)] $6\lesssim\eta\lesssim 50$. The smoking ring gets
stretched, and the elementary links becomes aligned over long
distance. Fractal dimension shrinks to one, with the apparition of a
long range order. Rotational invariance is broken.

\item[(d)] $\eta\gtrsim50$. The system is not able to reach its
equilibrium. This typical configuration (out of equilibrium) exhibits
an intricate hierarchical structure of bubbles. Smaller bubbles get
evaporated into bigger bubbles. Time constants associated to these
mechanisms grow exponentially with the size of the bubbles. Therefore,
the final configuration corresponding to one bubble is never reached
in the thermodynamic limit. We will refer to this non-equilibrium
phase as a \emph{glassy} phase.
\end{itemize}

\subsection{Brownian windings}

Some macroscopic random variables of interest can be constructed in
order to be able to follow numerically the evolution of the system.
First of all the total number of patterns $M1, M2, M3, M4$ is a set a
variables which can be used to distinguish between a folded and a
stretched phase of the system. All these number are expected to be
fairly distributed in the folded phase although in the stretched phase
we expect the pattern $M4$ to be in majority. In order to express
mathematically the curling of the system, we consider variables
related to the winding properties of planar Brownian curves, which are
defined as follows: with each point in the plane is associated its
\emph{winding angle} $\theta$, scanned by the random walker around
this point.
\begin{figure}[htb]
\begin{center}
\begin{tabular}{cc}
\resizebox*{!}{3.5cm}{ \input{wind.pstex_t}}&\hspace{.5cm}
\includegraphics[width=0.4\textwidth]{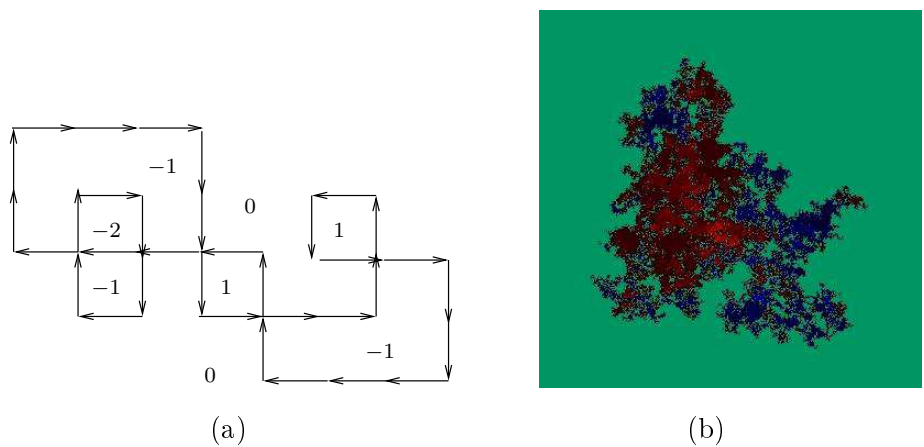}\\[0.2cm]
(a)&(b)
\end{tabular}
\caption{Winding sectors defined for a closed random walk (a). Color
gradient representing sectors for a walk of $N=10^6$ steps: in red and
blue, respectively, positive and negative winding sectors; in green,
the null sector (b).}
\end{center}
\end{figure}
A limit law for this variable has been derived by Spitzer
for Brownian curves. Actually, assuming the length of the curve is set
to $2\ell= N a^2$ when $N\to\infty$ and $a\to 0$ in the Brownian limit,
the winding angle $\theta(\ell)$ of an arbitrary point has the asymptotic
probability \cite{SPITZER}
\[ \lim_{\ell\to\infty} P\left(\theta(\ell) =
\frac{\alpha\log \ell}{2}\right)=\frac{1}{\pi}\frac{1}{1+\alpha^2}.
\]
 For a closed walk, the value taken by $\theta$ are limited to
$2n\pi$, where $n \in {\mathbb Z}$ represents the \emph{winding
number} associated with the point under consideration.

For each $n$, the set of points with the same winding number $n$ form
a \emph{winding sector}, whose
arithmetic area is a random variable
denoted by $S_n(\ell)$.
\begin{figure}[htb]
\begin{center}
\hspace{-1.5cm}\begin{tabular}{cc}
\includegraphics[angle=-90, width=0.5\textwidth]{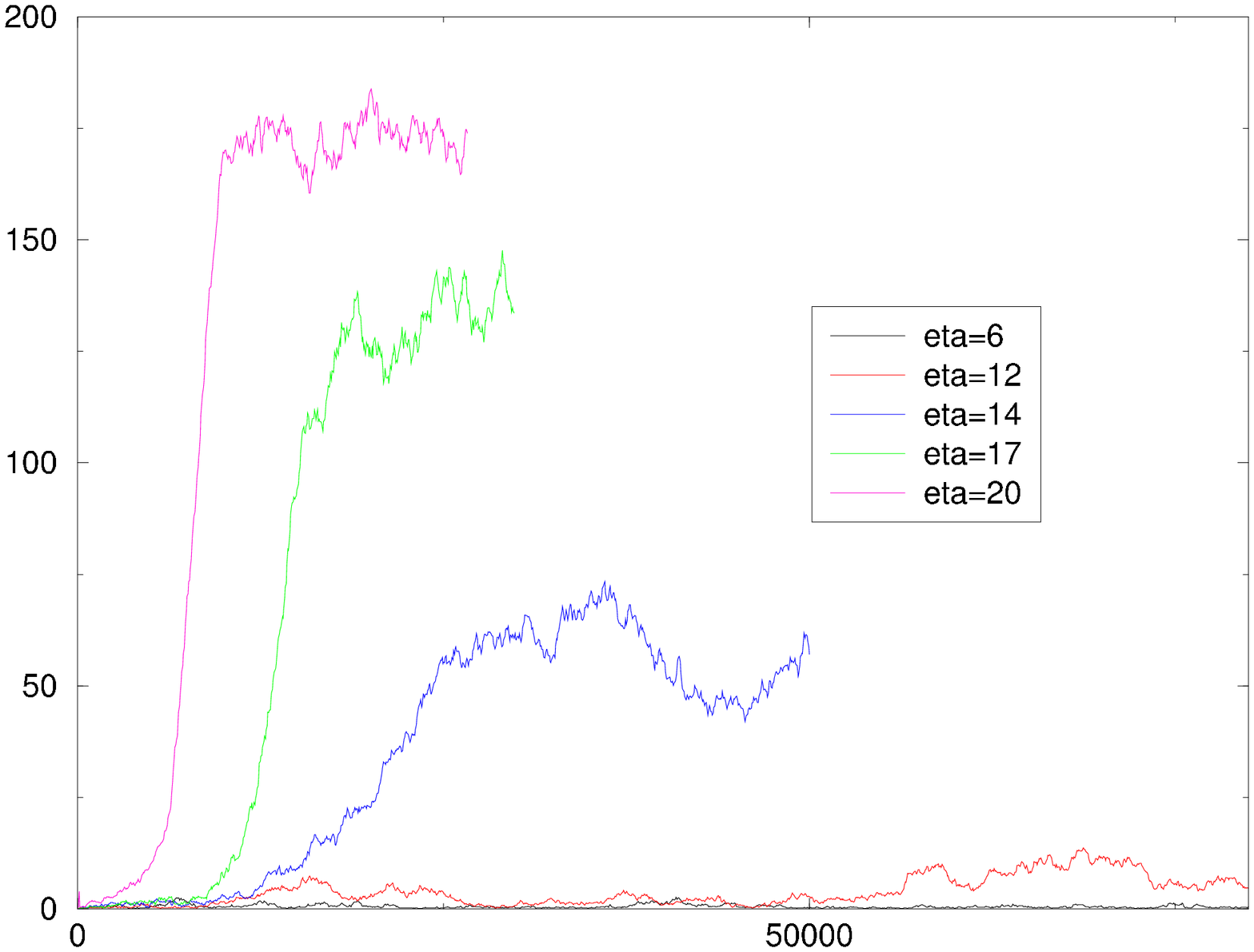}&
\put(0.2,.25){\includegraphics[angle=180, width=0.535\textwidth]{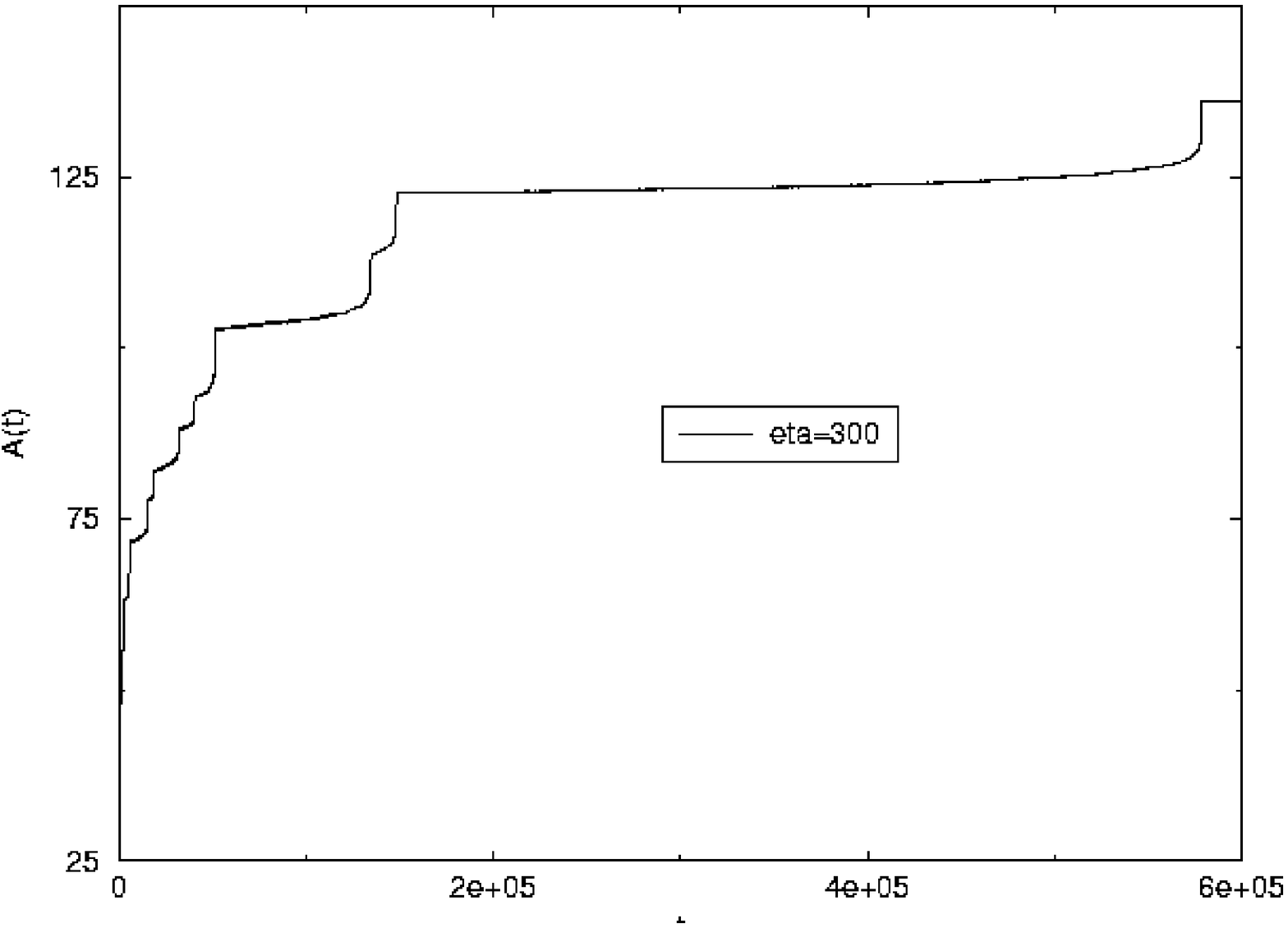}}\\
(a)&(b)\\[0.2cm]
\includegraphics[angle=-90, width=0.52\textwidth]{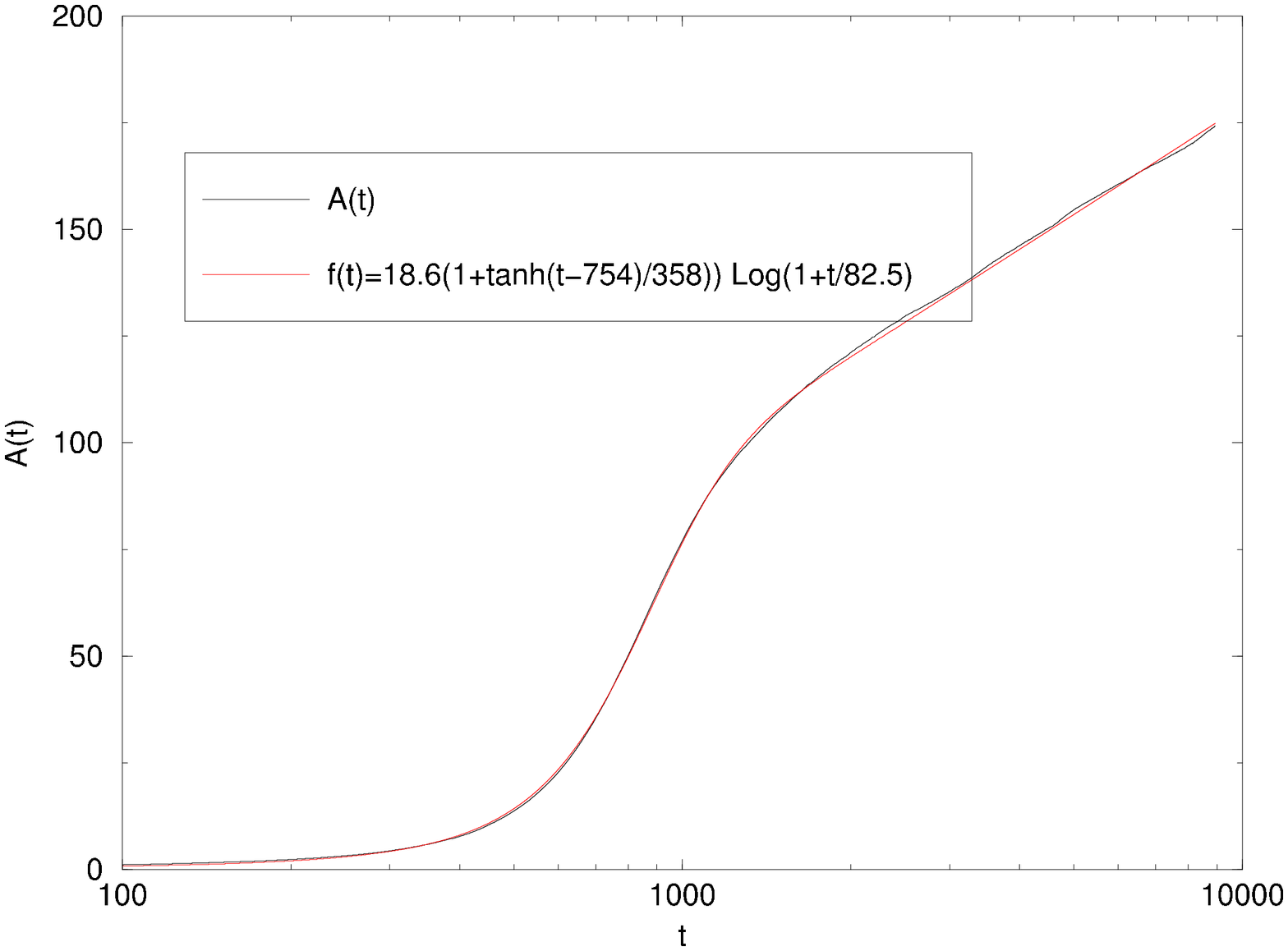}&
\includegraphics[angle=-90, width=0.51\textwidth]{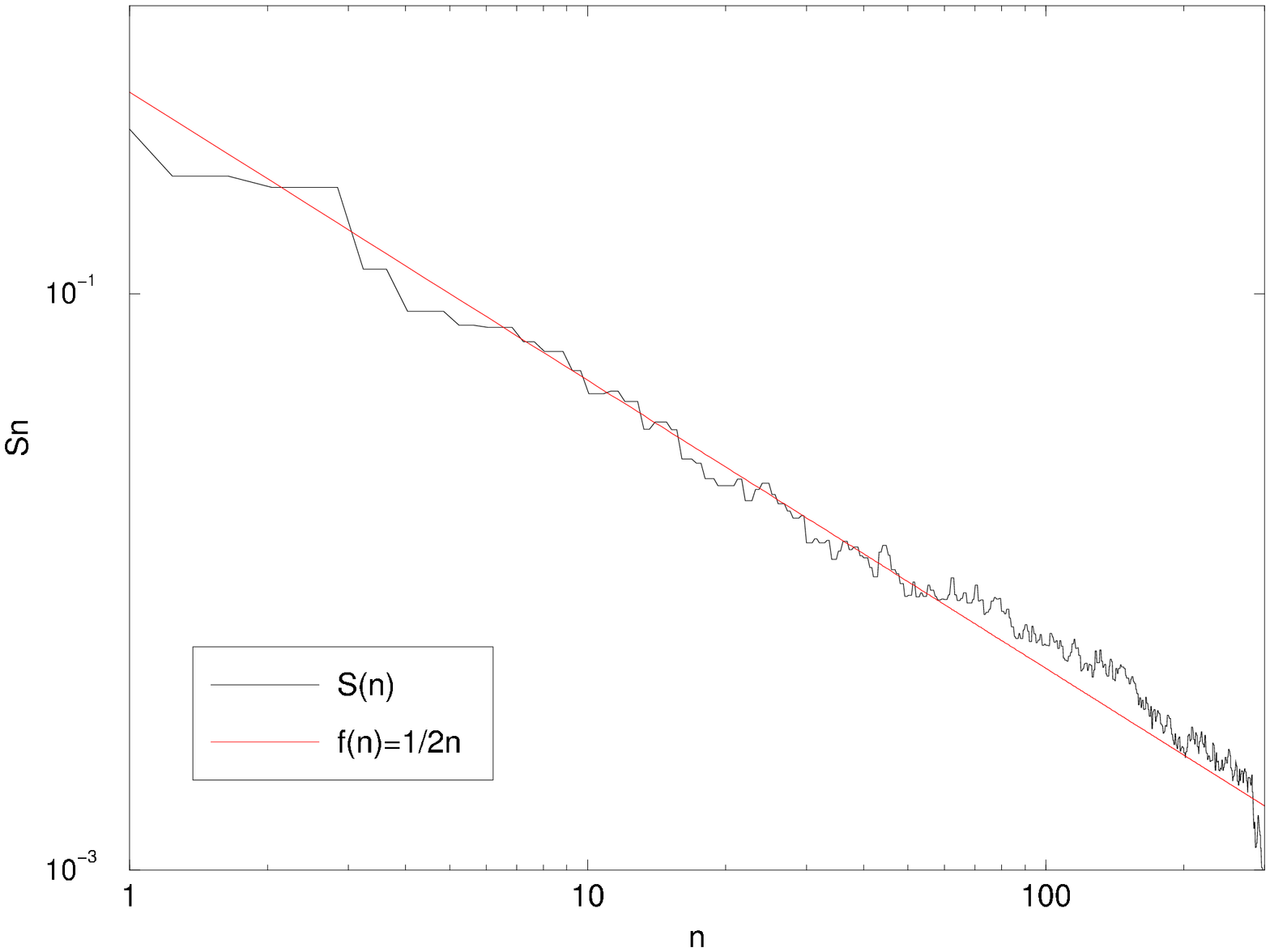}\\
(c)&(d)
\end{tabular}
\end{center}
\caption{\label{data}Evolution of the total algebraic area (a,b,c).
First transition with condensation in the first winding sector (a)
$N=1000$ steps. Unfolding transition with metastable states for
$N=2000$ steps (b). Curves (c) and (d), drawn in
a logarithmic scale for $N=10^6$, represent respectively the slow dynamics 
and the distribution of the $S_n$'s.
}
\end{figure} 
Under the above mentioned Brownian limit for closed
curves of length $\ell$, we have
\cite{COMTET}
\[
\mathbb{E}\left[S_n(\ell)\right] =
\frac{\ell}{2\pi n^2}.
\] 
In a similar way, the total \emph{algebraic
area} enclosed by the Brownian curve is defined as
\[ A(\ell)=\sum_{n\in\mathbb{Z}} nS_n(\ell) ,
\] 
and its distribution is given by L\'evy's
law \cite{LEVY} 
\[ 
\lim_{\ell\to\infty} P\left(A(\ell) =2s\ell\right) =
\frac{\pi}{2\cosh^2(\pi s)}. 
\] 
The variable $S_n(\ell)$ is indeed enough for a complete characterization
 of the phases of the system.
\begin{figure}[htb]
\hspace{-0.5cm}\begin{tabular}{cc}
\includegraphics[angle=-90, width=0.5\textwidth]{trans.eps}&
\includegraphics[angle=-90, width=0.5\textwidth]{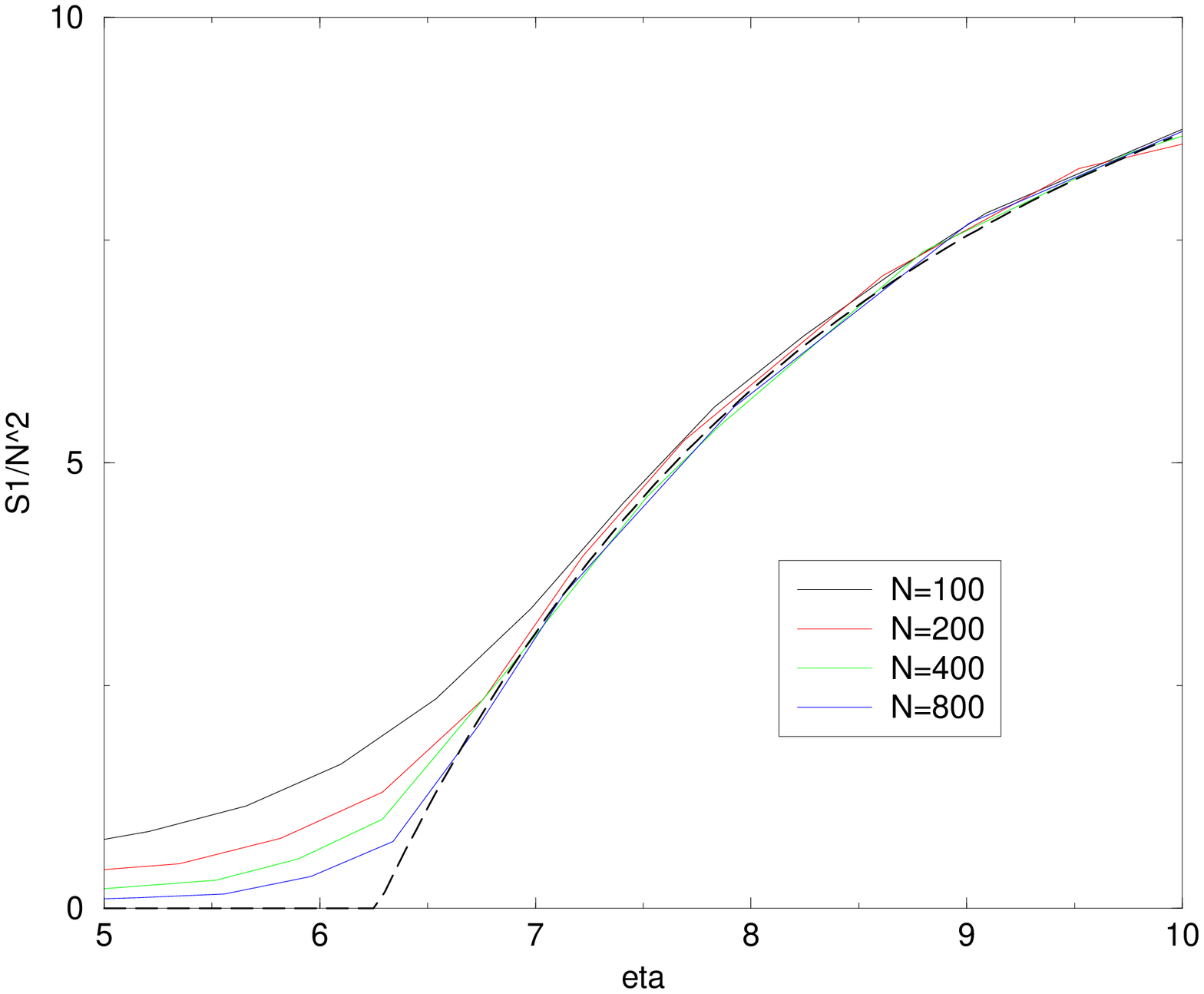}\\
(a)&(b)\\
\vspace{5cm}
\begin{picture}(0,0)
\put(1,0){\includegraphics[angle=-90, width=0.4\textwidth]{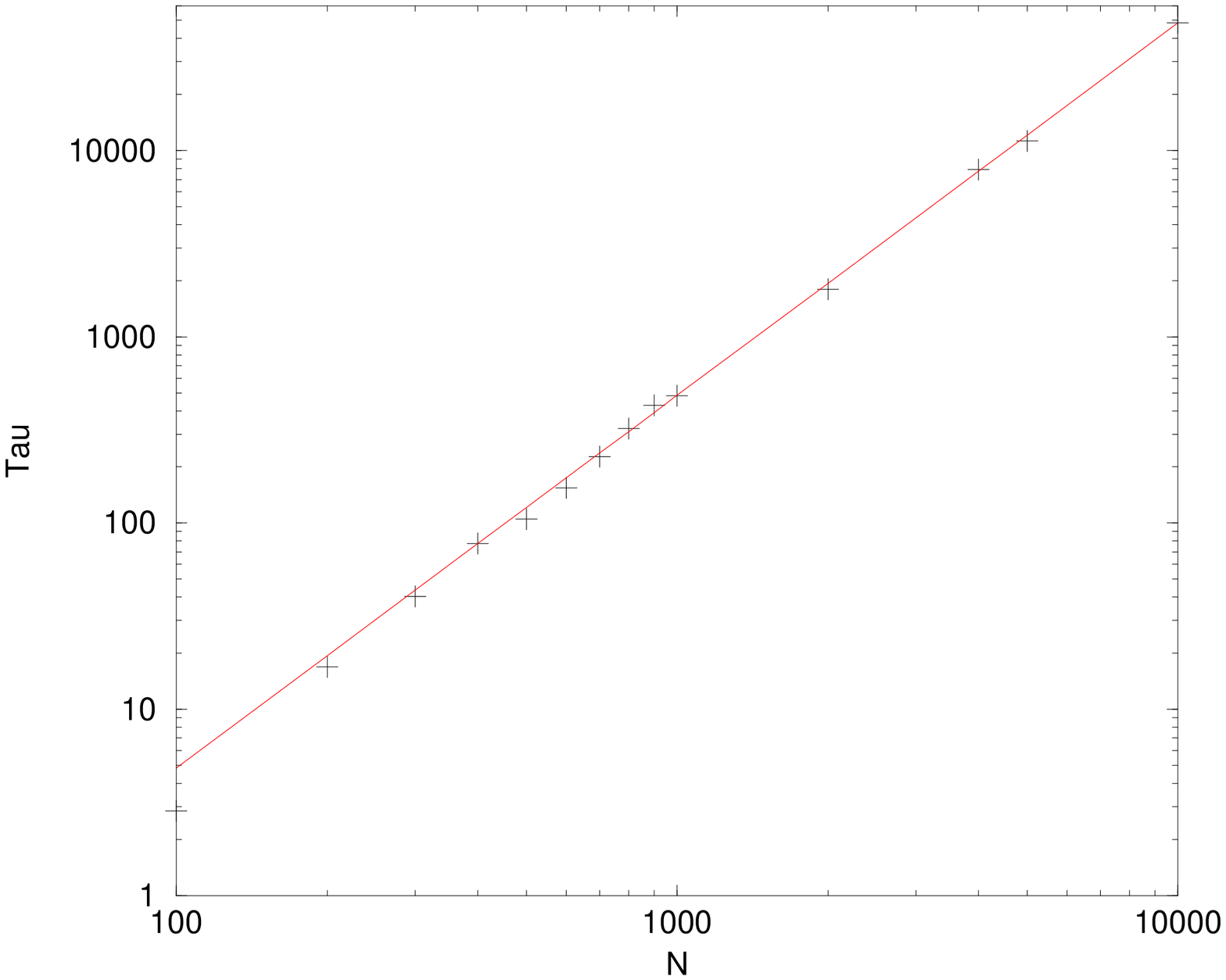}}
\put(3.5,-4.5){(c)}
\end{picture}\\
\end{tabular}
\caption{\label{etac} The limit arithmetic area
of the first winding sector when $\eta$ varies,  rescaled by $N$ (a) and by $N^2$ (b).
The dashed line in (b) is computed from part \ref{SECW} giving 
the critical 
value $\eta_c=2\pi$ separating the folded phases from stretched one. 
Scaling with $N$ of 
time-constants for the transient regime (c) for $\eta=10$; the fit gives a value 
$2.01$ for the exponent. }
\end{figure} 

\subsection{Slow dynamics}

The study of winding sectors and of associated variables is especially
well adapted to describe the evolution of the system, since the
presence of curls has a direct impact on the $S_n$ distribution. A
small but finite value of $\eta\lesssim 1$ will show itself by the
existence of a shift in the distribution of $S_n$ together with an
unbalancing between positive and negative winding sectors. When
$1\lesssim\eta\lesssim 6$, the distribution of $S_n$ condenses into
either $S_1$ or $S_{-1}$. However, the behavior of $S_1$ or $S_{-1}$
with regard to scaling factor does not change, and these variables
still scale like $N$. The system reaches its stationary state after a
transient regime characterized by a single time constant
(Figure~\ref{data}a). For $6\lesssim\eta\lesssim 50$ the system get
stretched, an unfolding transition occurs. $S_1$ or $S_{-1}$ scale now
like $N^2$, and the fact that the walk is taut get reflected in the
distribution of the motifs $M_1,M_2,M_3,M_4$. When $\eta\gtrsim50$ we
obtained a glassy phase. This is related to the apparition of a
hierarchy distribution of meta-stable configurations. The system
evolves slowly to the rate of bubbles evaporations, small bubbles
collapses and produce bubbles of bigger size. As a consequence, the
transitory regime is completely different. For small size systems (see
Figure~\ref{data}b), it is observed that the total algebraic area
increases by successive steps. These steps correspond to intermediate
metastable states, consisting of bubbles of increasing size. The
associated time constants behave roughly exponentially with the size
of these bubbles. When $N\to\infty$ (see Figure~\ref{data}c), a
continuous spectrum of time constants is obtained and the convolution
of these dynamical effect corresponding to different scale end up in a
slow dynamical grows. The total algebraic area increases
logarithmically with time. We observe also (see Figure~\ref{data}d)
that the distribution of $S_n$ seems to have a limit characterized by
the absence of negative (or positive) sectors, i.e. strictly zero for
negative (or positive) index $n$, together with a scaling exponent
around $1$. Indeed this sequence of distributions seems to behave like
$n^{-1}$, instead of $n^{-2}$ for $\eta=0$. Also the
glassy transition is clearly of first order, with coexistence of a
liquid phase (part of the walk which remains folded and disordered)
and a solid glassy phase. The parameters corresponding to temperature
and magnetization can be defined by analogy with standard spin
glasses. They can be tuned independently, as we shall see later on, by
letting vary $\eta$ (which is roughly equivalent to the external
magnetic field) and $\gamma$ (pertaining in some sense to the
temperature).

\begin{figure}[htb]
\begin{center}
\begin{tabular}{cc}
\includegraphics[width=0.4\textwidth]{f6.eps}&
\hspace{0.5cm}\includegraphics[width=0.4\textwidth]{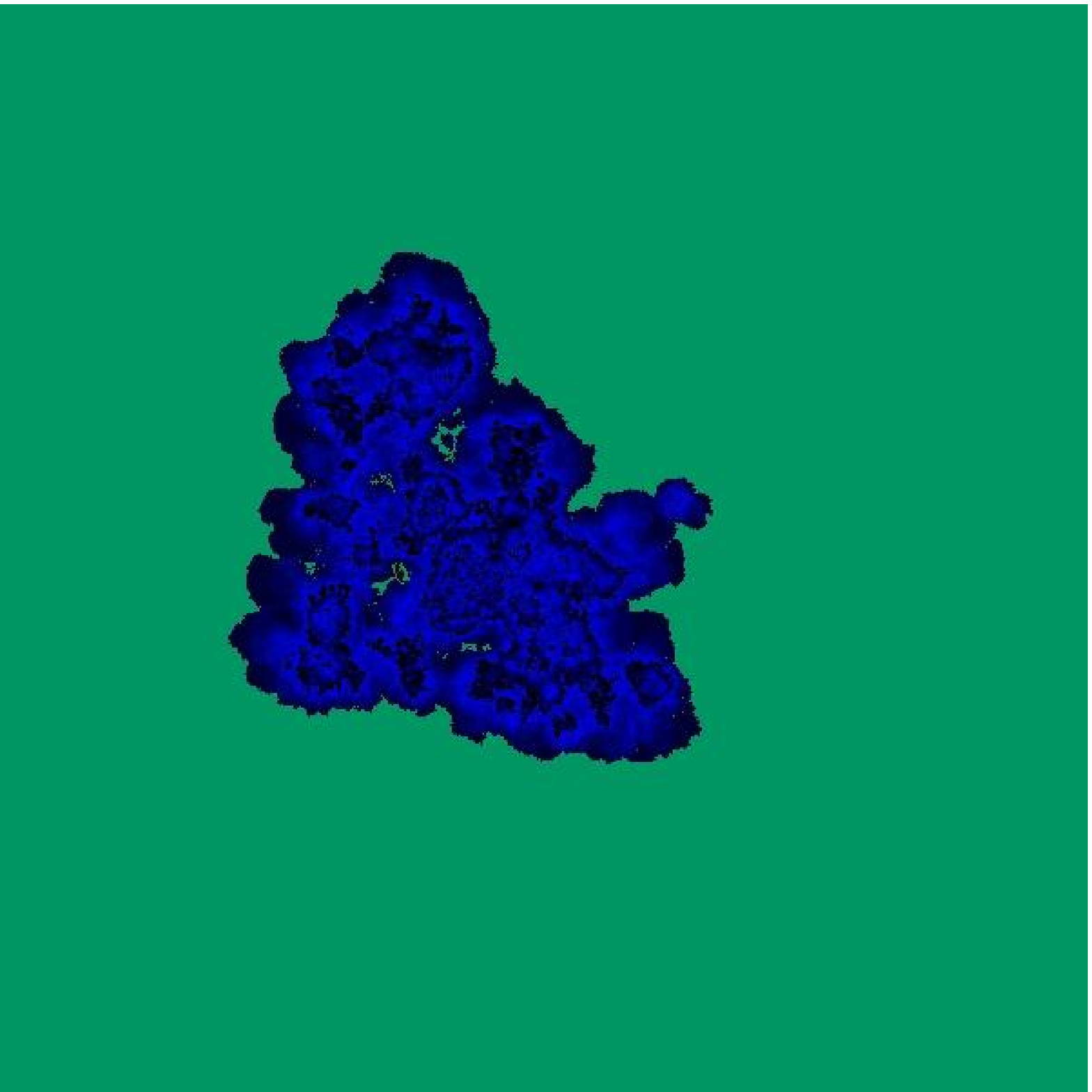}\\
(a)&(b)\\[0.2cm]
\includegraphics[width=0.4\textwidth]{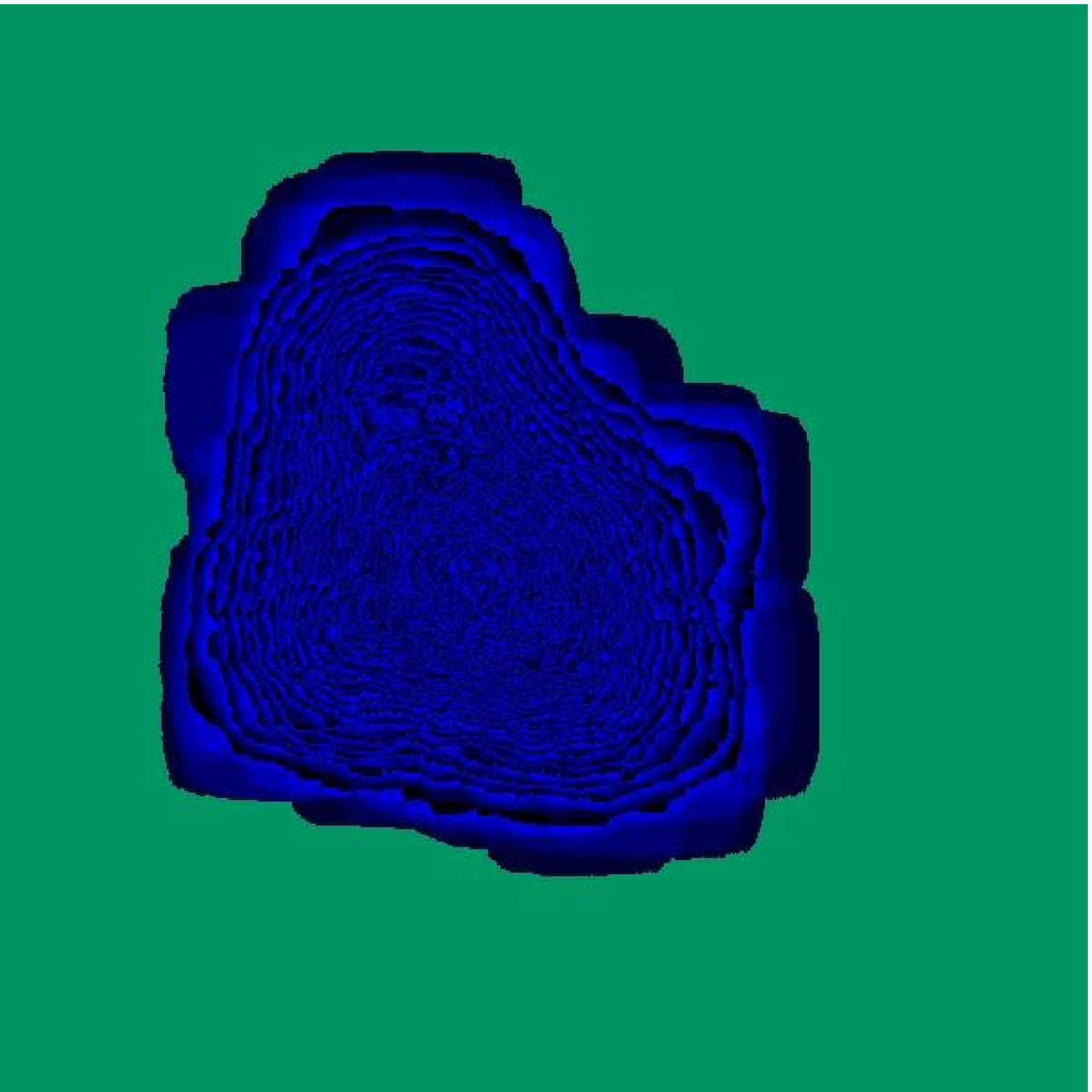}&
\hspace{0.5cm}\includegraphics[width=0.4\textwidth]{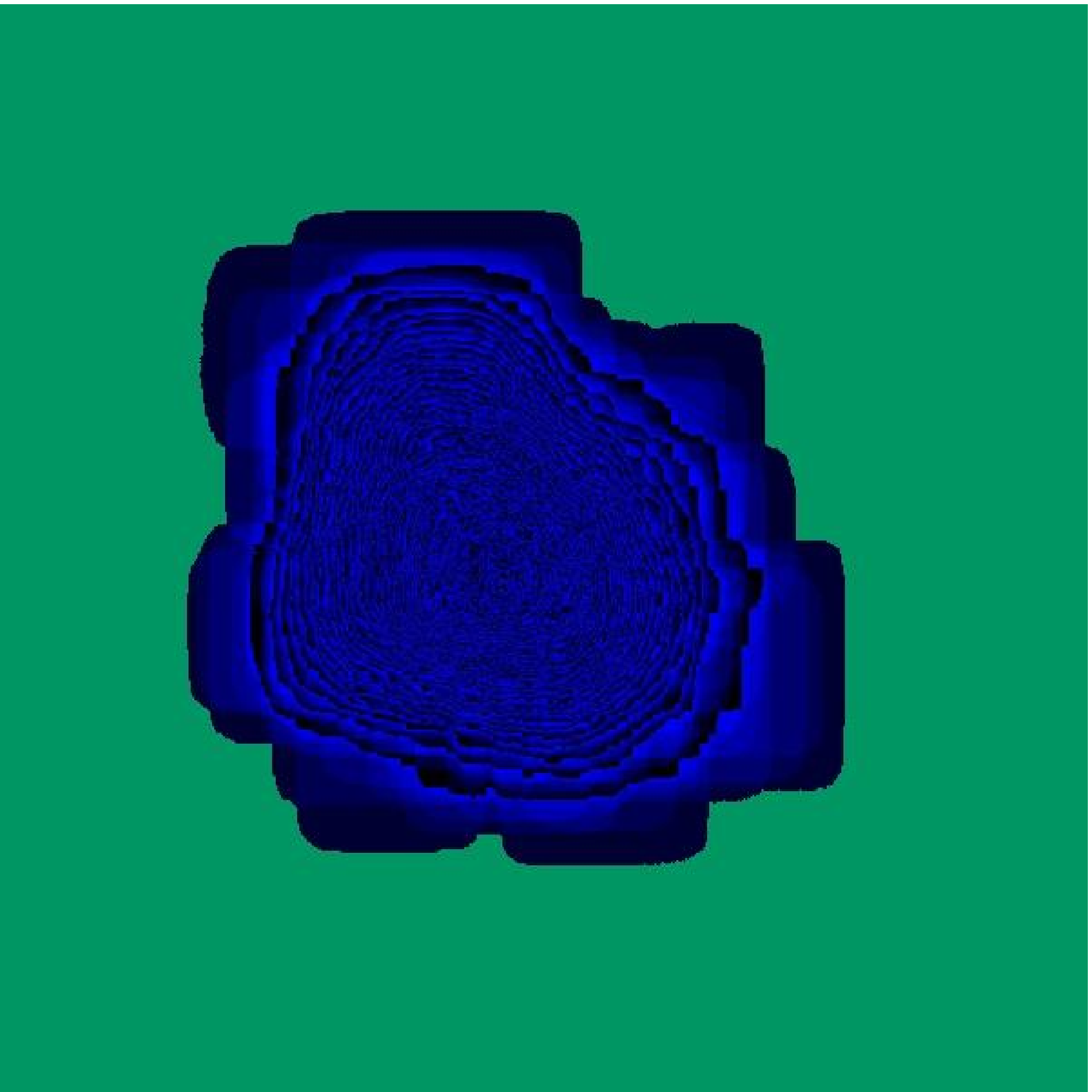}\\
(c)&(d)
\end{tabular}\caption{First order glassy transition, with coexistence of a solid
phase (on the edges) and a liquid phase (at the center). The scale is
divided by a factor of 2 in (c) and of 4 in (d). The blue gradient
indicates that all  windings are negatives. A complete sequence of
blue corresponds to an increase of 12 winding numbers.}
\end{center}
\end{figure}
\section{\bf Sequence coding and  generators: two formal approaches }

\subsection{\bf A group decomposition}

The dynamics of the system is Markovian. Its description involves a
configuration space $\{\alpha\}$, a probability measure $P_{\alpha}$
and the generator $G_{\alpha\beta}$ of the evolution operator. Since
the elementary transformations are local, and associated with  each point
of the random chain, $G$ may be expressed as a sum of operators
\[ G_{\alpha\beta}=\sum_{i=1}^N g_{\alpha\beta}(i),
\]
Here it is to be understood that each $g(i)$ operates on a real space, endowed 
with the base
$\{\omega_{\alpha_1}(1)\otimes\omega_{\alpha_2}(2)...\otimes\omega_{\alpha_N}(N)\}$
defined tensorially with respect to sites $i$, where each
$\alpha_i$ stands for possible configurations at position $i$.
Therefore, the sequence $\{\alpha_1,...,\alpha_N\}$ will generically
denote a given configuration of the system. Then $g$ has the following
representation in the base of local operators
\begin{eqnarray*}
g(i) &=&
\lambda^+\sigma^+(i)(1-\sigma^-(i))\ +\
\lambda^-\sigma^-(i)(1-\sigma^+(i)) \nonumber \\ &+& \gamma^+
(R(i)-R^4(i))\ +\ \gamma^-(R^{3}-R^4(i)) ,
\end{eqnarray*}
where $\sigma^+(i),\sigma^-(i),R(i),R^{3}(i)$ are the elementary
operators corresponding to the  transformations given in the
introduction. They enjoy the following properties:
\begin{eqnarray*} \sigma^{\pm}(i)^2 &=& 0 ,\\
\sigma^{\pm}(i)\sigma^{\mp}(i)\sigma^{\pm}(i) &=& \sigma^{\pm}(i) ,\\
\sigma^{\pm}(i)R(i) &=&R(i)\sigma^{\pm}(i) \ = \ 0 ,\\ R^5(i)&=&R(i).
\end{eqnarray*}
The fact that the transformations are local is illustrated by the
commutation property
\[ h(i)h(i+2)\ =\ h(i+2)h(i). 
\]
Many representations can be defined to encode system
configurations. For example, we could take the sequence
\[\Omega_0=\{z_1=x+iy, z_2,...,z_N\}, \ (z_i\in \mathbb{Z}+i\mathbb{Z})
\]
formed by the affixes of the successive points in the plane, with the
constraint that each point is separated from its neighbors by a
single link. Another possible choice is 
\[
\Omega_1=\{v_1=z_2-z_1,v_2=z_3-z_2,...,v_{N-1}=z_N-z_{N-1}\},\ (v_i
\in \{1,i,-1,-i\}), 
\]
the set of successive links, which corresponds to the
tangent map to $\Omega_0$. In this case, the center of gravity, i.e. the
average over all point positions $z^*=(z_1+..+z_N)/N$,
defines equivalent classes. Then taking  the tangent map to $\Omega_1$ yields
 the sequence
\[
\Omega_2=\{u_1=v_2-v_1,u_2,...,u_{N-2}=v_{N-1}-v_{N-2}\},
\]
which represents the patterns defined in introduction. In that case, the
drift 
\[v^*=(v_1+\cdots v_{N-1})/N
\]
 depicts the separation between extremities, and may serve to define
different equivalent classes. We will see later that, depending on the
boundary conditions , $z^*$ and $v^*$ can be decoupled from the
dynamics. In particular, this is true for periodic boundary
conditions. Indeed $\Omega_2$ should be the natural representation to
express the dynamics of the system, since elementary transformations
are performed on the patterns defined at each point of the chain.
However, although any possible transition at a given point depends
solely on the pattern of the point (for example a
$\lambda^+$-transition at point $i$ can take place only if the walk
performs a left bend at that point), the result of this transformation
involves three successive patterns (for the event $\lambda^+$, the
pattern at $i$ becomes a right bend, and both adjacent patterns at
$i-1$ and $i+1$ are modified). On the other hand, in the $\Omega_1$
representation, although each transition is conditioned by a pair of
links, the result does modify only this pair.

We choose now $\Omega_1$ to express $G$ properly. As a site $i$ of the
chain has four possible configurations, depending on the orientation
of the link $v_i$, let us introduce
$\{\omega_0(i),\omega_1(i),\omega_2(i),\omega_3(i)\}$ the
corresponding base in a $four$-dimensional space, and consider
$\phi^{\alpha,\beta}(i)$, $\alpha,\beta\in\{0,1,2,3\}$, the set of
operators defined on this space by
\[
\phi^{\alpha,\beta}\omega_\mu= \delta_{\beta\mu}\omega_\alpha. 
\] In
particular $\phi^{\alpha,\alpha}(i)$ is the projection onto the state
$\omega_{\alpha}$ at site $i$, the remainder of the sequence being
left unmodified.  With this notation, the generator takes the form
\bea\label{gdi} 
g(i)&=& \lambda^+\left[
\phi^{\alpha+1,\alpha}(i)\phi^{\alpha,\alpha+1}(i+1) -
\phi^{\alpha,\alpha}(i)\phi^{\alpha+1,\alpha+1}(i+1) \right] \nonumber
\\
&+&\lambda^-\left[ \phi^{\alpha,\alpha+1}(i)\phi^{\alpha+1,\alpha}(i+1) - 
\phi^{\alpha+1,\alpha+1}(i)\phi^{\alpha,\alpha}(i+1) \right] \nonumber \\
&+&\gamma^+\left[ \phi^{\alpha+1,\alpha}(i)\phi^{\alpha-1,\alpha}(i+1) - 
\phi^{\alpha,\alpha}(i)\phi^{\alpha+2,\alpha+2}(i+1) \right] \nonumber \\
&+&\gamma^-\left[\phi^{\alpha,\alpha+1}(i)\phi^{\alpha,\alpha-1}(i+1) - 
\phi^{\alpha,\alpha}(i)\phi^{\alpha+2,\alpha+2}(i+1) \right], 
\eea
and commutation rules write
\[
[\phi^{\alpha,\beta}(i),\phi^{\mu,\nu}(j)]= \delta_{ij}\left(
\delta_{\beta\mu}\phi^{\alpha,\nu}(i)-\delta_{\alpha\nu}\
\phi^{\mu\beta}(i)\right).
\]
The form (\ref{gdi}) could be used to construct a transfer matrix
representation \cite{HAKIM}.  For the moment, one can simply remark
that the Lie algebra $\cal L$ generated by the $\phi^{\mu\nu}$ is a
${\mathbb Z}_2$-graded Lie algebra \cite{NOVIKOV}, so that it can be
decomposed onto two Lie sub-algebras denoted by ${\cal L}_0$ and ${\cal L}_1$,
with the commutation relations
\[
\begin{cases} 
 \ [{\cal L}_0,{\cal L}_0]\subset {\cal L}_0 ,\\[0.2cm]  
 \ [{\cal L}_0,{\cal L}_1] \subset {\cal L}_1 ,\\[0.2cm]  
\ [{\cal L}_1,{\cal L}_1] \subset {\cal L}_0.
\end{cases}
\]
This leads to introduce the following linear combinations
\be\label{Charges} \kappa^{p,q} =
\sum_{\alpha\in\{0,1,2,3\}}e^{\frac{i\pi p\alpha}{2}}\phi^{\alpha+q,\alpha},
\end{equation} 
for $p,q \in\{0,1,2,3\}$, with the commutation
rules
\[ 
[\kappa^{p,q},\kappa^{p',q'}]= \left( e^{\frac{i\pi
pq'}{2}}-e^{\frac{i\pi p'q}{2}}\right) \kappa^{p+p',q+q'},
\] 
 where the sums $p+p'$ and $q+q'$ are taken modulo $4$.  One easily
 verifies that, when $q$ varies, $\kappa^{0,q}$ and $\kappa^{2,q}$
 generate ${\cal L}_0$, while $\kappa^{1,q}$ and $\kappa^{3,q}$
 generate ${\cal L}_1$.  In addition, from the very definition
 (\ref{Charges}), the quantities
\[
\begin{cases}
 \ Q_0 = \sum_j  \kappa^{0,0}(j) ,\\[0.2cm]
 \ Q_1 = \sum_j  \kappa^{1,0}(j) ,\\[0.2cm]
 \ Q_3 = \sum_j  \kappa^{3,0}(j) ,\\
\end{cases}
\]
do commute with the generator.  Later, by means of a more intuitive
representation, we will be able to interpret the above formalism in terms
of conservation of particles.

\subsection{Chaos game representation}\label{CHAOS}
\begin{figure}[htb]
\begin{center}
\resizebox*{!}{5cm}{ \input{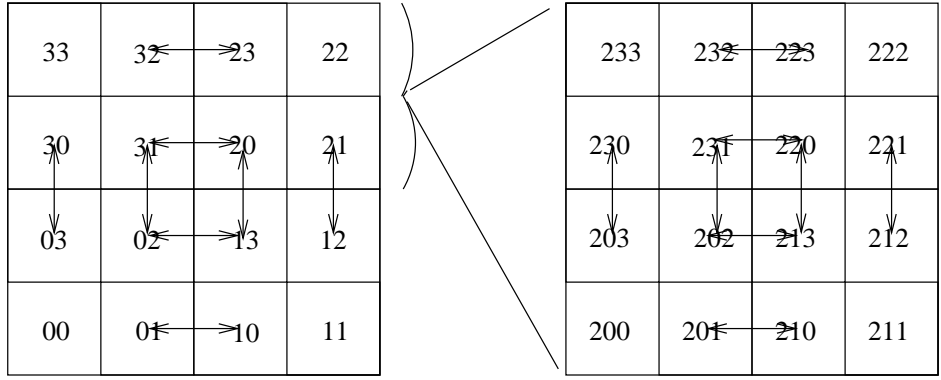}}
\caption{\label{quad}Chaos Game representation of the first two links
of the chain (left) and change of scale (right) representing the next
link. Arrows indicate the transitions $\gamma$ (center) and $\lambda$
(on the edges). }
\end{center}
\end{figure}

A way commonly used in  DNA modeling (\cite{DORFMAN,ALMEIDA} and
references therein) in order to visualize  dynamics in the
configuration space consists in encoding the random walk as an array
of integers $\alpha_i\in\{0,1,2,3\},\ i\in\mathbb{N}$, and to
associate with each sequence a complex number $z$, where 
\[ z=
\sum_{p=0}^{N-1}\frac{1}{2^p}\exp(i\pi\alpha_p/2). 
\]
Graphically, this leads to the successive inclusion of disjoint
square sectors  representing the filtration defined by the random walk
(see figure \ref{quad}).
To each random walk of length $N$ corresponds a square of area
$4^{-N}$ centered at the related point $z$.  There are exactly $4^N$
elementary squares and it becomes possible to define a discrete
probability measure $\pi(z,t)$, constant on each such square, and
whose evolution at time $t$ is given by the forward Kolmogorov's
equations
\[
\frac{\partial\pi}{\partial t}(z,t)=\int dy \pi(y,t)g(y,z).
\]
where $g(z,z')$ denotes the kernel of the infinitesimal generator in
the above representation.  Let $g_0$ be the part of the operator which
operates at the upper level of the representation, that is on the
first two links (see Figure~\ref{quad}), and let $D$ the dilation
operator mapping $z$ onto $2z$.  Then $g$ writes 
\[
g=g_0+Dg_0D^{-1}+D^2g_0D^{-2}+ \ldots
\] 
Moreover, as $N\to\infty$, $g$ must satisfy the consistency relation
\[ 
g=g_0+DgD^{-1},
\]
which can be split into two parts
 \[
 g\egaldef g_1+g_2, 
\]
where
\[
\begin{cases}
 g1=g_0+D^2g_1D^{-2} ,\\[0.2cm]
 g2=Dg_0D^{-1}+D^2g_2D^{-2}.
\end{cases}
\]
It appears that $g_1, g_2$ can be easily diagonalized, since $g_0$ is
diagonalizable and commutes with $D^2g_0D^{-2}$ and all higher order
terms. Hence $g_1, g_2$ can be endowed with a simple tensorial
structure which, in matrix form, reads

\begin{picture}(0,0)%
\includegraphics{g0.pstex}%
\end{picture}%
\setlength{\unitlength}{3947sp}%
\begingroup\makeatletter\ifx\SetFigFont\undefined%
\gdef\SetFigFont#1#2#3#4#5{%
  \reset@font\fontsize{#1}{#2pt}%
  \fontfamily{#3}\fontseries{#4}\fontshape{#5}%
  \selectfont}%
\fi\endgroup%
\begin{picture}(3876,3085)(2551,-4077)
\put(3766,-1681){\makebox(0,0)[lb]{\smash{\SetFigFont{7}{8.4}{\rmdefault}{\mddefault}{\updefault}$[-\gamma]$}}}
\put(4531,-2401){\makebox(0,0)[lb]{\smash{\SetFigFont{7}{8.4}{\rmdefault}{\mddefault}{\updefault}$[-\gamma]$}}}
\put(4666,-2580){\makebox(0,0)[lb]{\smash{\SetFigFont{7}{8.4}{\rmdefault}{\mddefault}{\updefault}$[-\lambda^-]$}}}
\put(4846,-2760){\makebox(0,0)[lb]{\smash{\SetFigFont{7}{8.4}{\rmdefault}{\mddefault}{\updefault}$[0]$}}}
\put(5611,-3480){\makebox(0,0)[lb]{\smash{\SetFigFont{7}{8.4}{\rmdefault}{\mddefault}{\updefault}$[0]$}}}
\put(5746,-3660){\makebox(0,0)[lb]{\smash{\SetFigFont{7}{8.4}{\rmdefault}{\mddefault}{\updefault}$[-\lambda^+]$}}}
\put(5926,-3840){\makebox(0,0)[lb]{\smash{\SetFigFont{7}{8.4}{\rmdefault}{\mddefault}{\updefault}$[-\gamma]$}}}
\put(6106,-4020){\makebox(0,0)[lb]{\smash{\SetFigFont{7}{8.4}{\rmdefault}{\mddefault}{\updefault}$[-\lambda^-]$}}}
\put(5791,-1860){\makebox(0,0)[lb]{\smash{\SetFigFont{7}{8.4}{\rmdefault}{\mddefault}{\updefault}$[\lambda^+]$}}}
\put(4306,-3300){\makebox(0,0)[lb]{\smash{\SetFigFont{7}{8.4}{\rmdefault}{\mddefault}{\updefault}$[\lambda^+]$}}}
\put(5026,-4020){\makebox(0,0)[lb]{\smash{\SetFigFont{7}{8.4}{\rmdefault}{\mddefault}{\updefault}$[\lambda^+]$}}}
\put(3766,-2401){\makebox(0,0)[lb]{\smash{\SetFigFont{7}{8.4}{\rmdefault}{\mddefault}{\updefault}$[\gamma^+]$}}}
\put(4531,-3120){\makebox(0,0)[lb]{\smash{\SetFigFont{7}{8.4}{\rmdefault}{\mddefault}{\updefault}$[\gamma^+]$}}}
\put(5206,-3840){\makebox(0,0)[lb]{\smash{\SetFigFont{7}{8.4}{\rmdefault}{\mddefault}{\updefault}$[\gamma^+]$}}}
\put(4531,-1681){\makebox(0,0)[lb]{\smash{\SetFigFont{7}{8.4}{\rmdefault}{\mddefault}{\updefault}$[\gamma^-]$}}}
\put(5431,-2220){\makebox(0,0)[lb]{\smash{\SetFigFont{7}{8.4}{\rmdefault}{\mddefault}{\updefault}$[\lambda^-]$}}}
\put(5206,-2401){\makebox(0,0)[lb]{\smash{\SetFigFont{7}{8.4}{\rmdefault}{\mddefault}{\updefault}$[\gamma^-]$}}}
\put(3766,-3840){\makebox(0,0)[lb]{\smash{\SetFigFont{7}{8.4}{\rmdefault}{\mddefault}{\updefault}$[\gamma^-]$}}}
\put(3451,-1321){\makebox(0,0)[lb]{\smash{\SetFigFont{7}{8.4}{\rmdefault}{\mddefault}{\updefault}$[0]$}}}
\put(4711,-1500){\makebox(0,0)[lb]{\smash{\SetFigFont{7}{8.4}{\rmdefault}{\mddefault}{\updefault}$[\lambda^-]$}}}
\put(4126,-2041){\makebox(0,0)[lb]{\smash{\SetFigFont{7}{8.4}{\rmdefault}{\mddefault}{\updefault}$[0]$}}}
\put(4981,-2940){\makebox(0,0)[lb]{\smash{\SetFigFont{7}{8.4}{\rmdefault}{\mddefault}{\updefault}$[-\lambda^+]$}}}
\put(5386,-3300){\makebox(0,0)[lb]{\smash{\SetFigFont{7}{8.4}{\rmdefault}{\mddefault}{\updefault}$[-\lambda^-]$}}}
\put(3991,-3660){\makebox(0,0)[lb]{\smash{\SetFigFont{7}{8.4}{\rmdefault}{\mddefault}{\updefault}$[\lambda^-]$}}}
\put(3586,-2580){\makebox(0,0)[lb]{\smash{\SetFigFont{7}{8.4}{\rmdefault}{\mddefault}{\updefault}$[\lambda^+]$}}}
\put(2551,-2671){\makebox(0,0)[lb]{\smash{\SetFigFont{12}{14.4}{\rmdefault}{\mddefault}{\updefault}$P_0\ =$}}}
\put(5971,-1681){\makebox(0,0)[lb]{\smash{\SetFigFont{7}{8.4}{\rmdefault}{\mddefault}{\updefault}$[\gamma^+]$}}}
\put(3586,-1500){\makebox(0,0)[lb]{\smash{\SetFigFont{7}{8.4}{\rmdefault}{\mddefault}{\updefault}$[-\lambda^+]$}}}
\put(3946,-1860){\makebox(0,0)[lb]{\smash{\SetFigFont{7}{8.4}{\rmdefault}{\mddefault}{\updefault}$[-\lambda^-]$}}}
\put(4261,-2220){\makebox(0,0)[lb]{\smash{\SetFigFont{7}{8.4}{\rmdefault}{\mddefault}{\updefault}$[-\lambda^+]$}}}
\put(5971,-3120){\makebox(0,0)[lb]{\smash{\SetFigFont{7}{8.4}{\rmdefault}{\mddefault}{\updefault}$[\gamma^-]$}}}
\put(6151,-2940){\makebox(0,0)[lb]{\smash{\SetFigFont{7}{8.4}{\rmdefault}{\mddefault}{\updefault}$[\lambda^-]$}}}
\put(5251,-3120){\makebox(0,0)[lb]{\smash{\SetFigFont{7}{8.4}{\rmdefault}{\mddefault}{\updefault}$[-\gamma]$}}}
\end{picture}

The self similarity is revealed by the fact that each $4\times4$ block
element becomes $16\times16$ when one takes an additional link, in
such a way that $P_0$ is added to itself in these blocks.  The problem
is intricate since $g_1$ and $g_2$ do not commute.  When
$\gamma^+=\gamma^-$, $P_0$ can be diagonalized and its eigenvalues are
$0$ (9 times degenerated), $-(\lambda^++\lambda^-)$ (4 times
degenerated), $-\gamma$ (twice degenerated) and $-2\gamma$ (simple). 
These simple tensorial structure simply reflects the scale invariance
in the chaos game representation or the translation invariance in the
initial representation.  When $N\to\infty$, it should be possible to
exploit this symmetry to build an iterative map, aiming at generating
the invariant measure or the self-organized critical one in the glassy
phase.  Before doing this, we will propose another formulation of the
problem in terms of particle hops with exclusion.

\section{Coupled exclusion models and thermodynamic limit}
\subsection{\label{Mapping}The mapping}
\begin{figure}[htb]
\vspace{0.5cm}
\begin{center}
\resizebox*{!}{8cm}{ \input{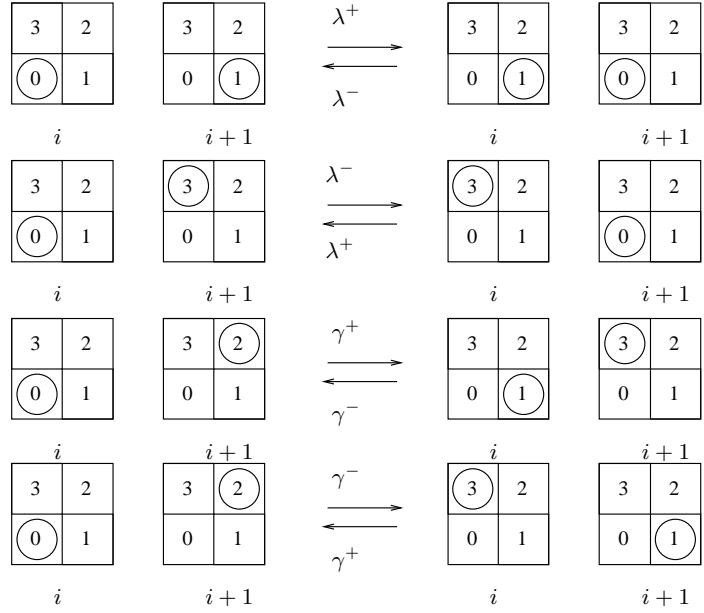}}
\caption{\label{bidi}Horizontal and vertical exchanges corresponding
to transitions when link $i$ has the value $0$, according to the
coding of section \ref{CHAOS}. Rules for other transitions follow
by rotational symmetry.}
\end{center}
\end{figure}

In the last section, we used a representation helping to visualize the
configuration space. There transitions between configurations were
expressed as exchanges between square domains. We may go a bit farther on,
using the fact that mutations are always either vertical or
horizontal, and with opposite directions between both modified links
(see Figure~\ref{bidi}). This suggests indeed to recode each link $j$,
$j=1,\ldots,N$ by means of two binary components $s_j^a\in\{0,1\}$ and
$s_j^b\in\{0,1\}$ thus establishing a \emph{bijection} $\alpha_j \to
(s_j^a,s_j^b)$ such that
\[
\begin{cases}
0 \to (0,0),\\
1 \to (1,0),\\
2 \to (1,1),\\
3 \to (0,1).\\
\end{cases}
\]
In this scheme a transition on a link does touch only either of its
components. Indeed the infinitesimal generator is the sum of two
terms: the first one acts on the sequence $\{s_i^a\}$, with rates
conditioned by the sequence $\{s_i^b\}$ and vice-versa. We have thus a
Markov process with state space
\[(S^a,S^b)=\left((s_1^a,s_1^b),...,(s_N^a,s_N^b)\right),
\] a
$2N$-dimensional boolean vector. Then, for an arbitrary function
$f:\,(S^a,S^b)\to \mathbb{C}$, the generator decomposes into
\be\label{G}
G=\sum_{i=1}^N h_a(i)+h_b(i),
\end{equation} 
where
\[
\begin{split}
h_a(i)f(S^a,S^b) &\egaldef
\lambda_a^+(i)s_i^a{\bar s}_{i+1}^a
\bigl[f\left((s_1^a,s_1^b),...,(0,s_i^b),(1,s_{i+1}^b),...,(s_N^a,s_N^b)\right)
-f(S^a,S^b)\bigr] \\
&+ \lambda_a^-(i){\bar s}_i^a s_{i+1}^a
\bigl[f\left((s_1^a,s_1^b),...,(1,s_i^b),(0,s_{i+1}^b),...,(s_N^a,s_N^b)\right)
- f(S^a,S^b)\bigr], \\[0.2cm]
h_b(i)f(S^a,S^b) &\egaldef
\lambda_b^+(i)s_i^b{\bar s}_{i+1}^b
\bigl[f\left((s_1^a,s_1^b),...,(s_i^a,0),(s_{i+1}^a,1),...,(s_N^a,s_N^b)\right)
-f(S^a,S^b)\bigr] \\ 
&+
\lambda_a^-(i){\bar s}_i^b s_{i+1}^b
\bigl[f\left((s_1^a,s_1^b),...,(s_i^a,1),(s_{i+1}^a,0),...,(s_N^a,s_N^b)\right)
-f(S^a,S^b)\bigr], 
\end{split}
\]
using the boolean notation ${\bar s}=1-s$. In order to express these
generators in terms of local operators, let us define the following
pairs of operators ($a_j$,$a_j^{\dag}$) and ($b_j$,$b_j^{\dag}$):
they leave unchanged the sample path, but the first [resp.
the second] component of link $j$, and they are given by
\bea\label{OPER} a_j f(S^a,S^b) &\egaldef& s_j^a
f\left((s_1^a,s_1^b),...,(0,s_j^b),...,(s_N^a,s_N^b)\right), \nonumber
\\ b_j f(S^a,S^b) &\egaldef& s_j^b
f\left((s_1^a,s_1^b),...,(s_j^a,0),...,(s_N^a,s_N^b)\right), \nonumber
\\ a_j^\dag f(S^a,S^b) &\egaldef& {\bar s}_j^a
f\left((s_1^a,s_1^b),...,(s_j^a,1),...,(s_N^a,s_N^b)\right), \nonumber
\\ b_j^\dag f(S^a,S^b) &\egaldef& {\bar s}_j^b
f\left((s_1^a,s_1^b),...,(1,s_j^b),...,(s_N^a,s_N^b)\right). \eea
In terms of these operators, 
we also have (for (a) species)
\[
h_a(i) =
\lambda_a^-(i)(1-a_ia_{i+1}^{\dag})a_i^{\dag} a_{i+1} +
\lambda_a^+(i)(1-a_{i+1} a_i^{\dag}) a_{i+1}^{\dag} a_i. 
\]
Transition rates can be computed by inspecting the different
cases, which yields the following expressions. 
\begin{equation}\label{taux}
\begin{cases}
\lambda_a^{\pm}(i)={\bar s_i^b}{\bar s_{i+1}^b}\lambda^{\mp}+s_i^b
s_{i+1}^b\lambda^{\pm} +
{\bar s_i^b}s_{i+1}^b\gamma^{\mp} + s_i^b{\bar s_{i+1}^b}\gamma^{\pm},\\[0.2cm]
\lambda_b^{\pm}(i)={\bar s_i^a}{\bar s_{i+1}^a}\lambda^{\pm}+s_i^a
s_{i+1}^a\lambda^{\mp} +
{\bar s_i^a}s_{i+1}^a\gamma^{\pm} + s_i^a{\bar s_{i+1}^a}\gamma^{\mp}.
\end{cases}
\end{equation}

The generator (\ref{G}) represents two coupled systems of particles
moving on a $one$-dimensional lattice with exclusion (i.e. there is at most
one particle of each species per site \cite{LIG}).
\begin{figure}[htb]
\begin{center}
\resizebox*{!}{8cm}{ \input{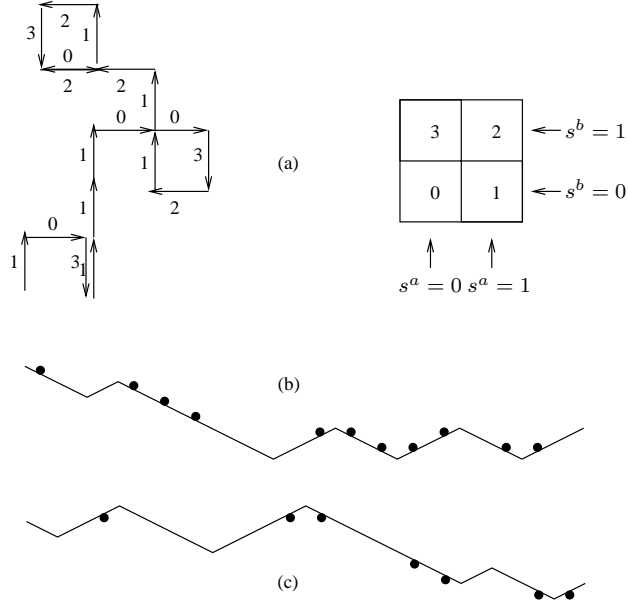}}
\caption{\label{MAP}Correspondence between chaos game representation
of the random walk (a) and one-dimensional models of particle
diffusion with exclusion. In (b), the sequence $\{s^a\}$ [resp.
$\{s^b\}$] determines the profile of the diffusion [resp. the
distribution of particles], drawn for $\gamma^{\pm}=\lambda^{\pm}$;
in figure (c) the role of the particles has been exchanged.}
\end{center}
\end{figure}

These particles perform random elementary jumps to the left or to the
 right. Obviously, both systems are interlaced: the jump rates
 $\lambda_a^{\pm}(i)$ of species $(a)$ at site $i$ are conditioned by
 the states of particles of species $b$ at site $i,i+1$, and
 conversely according to relations (\ref{taux}). Setting 
\bea\label{LM}
 \lambda &\egaldef& \frac{\lambda^++\lambda^-}{2} \quad \textrm{and}
 \quad \mu \egaldef \frac{\lambda^+-\lambda^-}{2}, \nonumber \\ \gamma
 &\egaldef&\frac{\gamma^++\gamma^-}{2}\quad \textrm{and} \quad
 \delta\egaldef\frac{\gamma^+-\gamma^-}{2}, 
\eea
 (with similar definitions for $\lambda_a,\lambda_b$) the sub-generator of
 species (a) rewrites \bea h_a(i)&=&\lambda_a(i)
 (a_i^{\dag}-a_{i+1}^{\dag})(a_{i+1}-a_i)\nonumber \\ &
 &+\mu_a(i)\left[a_i^{\dag}(a_{i+1}-a_i)+(a_i^{\dag}-a_{i+1}^{\dag})a_i\right]
 + V_a(i),\nonumber \eea where $\lambda_a(i)$ and $\mu_a(i)$ denote
 respectively the diffusion coefficient and the drift (with the same
 notation as in (\ref{LM})). After a routine algebra, we obtain
 \bea\label{LBDMU} \lambda_a(i) &=& \lambda\ +\ (\gamma-\lambda)\left(
 s_{i+1}^b-s_i^b\right)^2,\nonumber \\ \mu_a(i)
 &=& \mu(1-s_i^b-s_{i+1}^b) + \delta(s_{i+1}^b-s_i^b),\nonumber\\
 \lambda_b(i) &=& \lambda\ +\ (\gamma-\lambda)\left(
 s_{i+1}^a-s_i^a\right)^2, \nonumber\\ \mu_b(i)
 &=& \mu(s_i^a+s_{i+1}^a-1) + \delta(s_i^a-s_{i+1}^a). \eea It is
 worth remarking $\lambda_a$ and $\lambda_b$ are strictly positive
 except when $\gamma$ equals zero, in which case they might be zero at
 certain points. The interaction term takes the form
\[ V_a(i)= 2\lambda_a(i)
s_i^a s_{i+1}^a + s_i^a(\mu_a(i)-\mu_a(i-1)),
\] 
where $\mu_a(i)$ can be interpreted as a potential. A convenient
representation of the system is to draw  a
one-dimensional profile from the sequence $\{s^b\}$ (positive or
negative slope depending on whether $s^b$ equals 0 or 1), to sketch
the probabilistic inclination to turn left or right, as shown in Figures
\ref{MAP}b,~\ref{MAP}c.

In the
particular case  $\gamma^{\pm}=\lambda^{\pm}$, we get
\begin{equation}\label{LAB} 
\begin{cases}
\lambda_a^{\pm}(i)=\frac{1}{2}
(\lambda^++\lambda^-)\pm(s_i^b-\frac{1}{2})(\lambda^+-\lambda^-) ,\\[0.3cm]
\lambda_b^{\pm}(i)=\frac{1}{2}(\lambda^++\lambda^-)
\mp(s_i^a-\frac{1}{2})(\lambda^+-\lambda^-) ,
\end{cases}
\end{equation}
and the sub-generator of species (a) takes the form
\[
h_a(i)=\lambda a_i^{\dag}\big( a_{i+1}+a_{i-1}-2a_i\big) +
\mu(2s_i^b-1)\big(a_i^{\dag}- a_{i+1}^{\dag}\big)\big(a_i+a_{i+1}\big)
+V_a(i), 
\] 
where $V_a(i)$ is a diagonal term,
\[ V_a(i)=\lambda s_i^a(s_{i+1}^a+s_{i-1}^a).
\]
In this particular case, $\lambda$ represents explicitly the diffusion
constant of an isolated particle, with drift $\mu$, the sign of which
is determined by $s_i^b$. $V_a(i)$ is reminiscent of the
interaction between particles coming from the exclusion constraint.

\begin{figure}[htb]
\begin{center}
\resizebox*{!}{2cm}{ 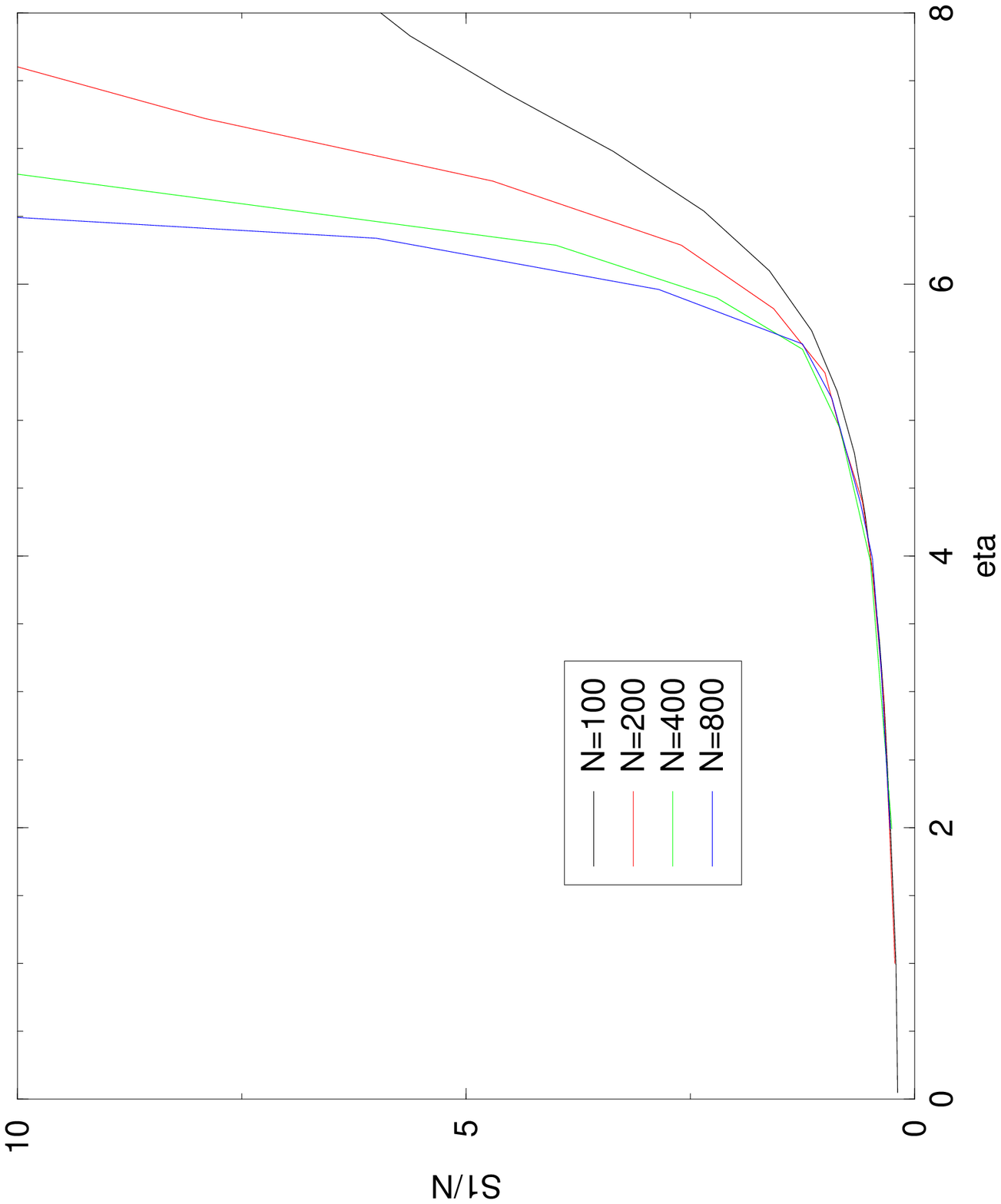}
\caption{\label{trans1}(a) Elementary transition due to the jump of a type $(a)$
particle, and the corresponding deformation of the profile defined by
$s^a$ (b) ($\gamma^{\pm}=\lambda^{\pm}$) }
\end{center}
\end{figure}
The distribution of particles labeled $(a)$, submitted to the
diffusion defined this way, is given by the sequence $\{s^a\}$ (0 or 1
particle depending on the individual site values of $s_a$). A
complementary model is obtained by exchanging the roles of $s^a$ and
$s^b$. In addition, elementary transitions of the system correspond to
jumps in the left or right direction of particles $(a)$ and $(b)$. In
the complementary formulation, these transitions are materialized
through modifications of the profile determined by $s^a$ or $s^b$ (see
figure \ref{trans1}). Therefore, from this viewpoint, we can formulate
the dynamic of subsystem (b) in terms of a KPZ model \cite{KPZ}, in
which the noise is produced by the distribution of particles (a). With
this formulation, the conserved quantities (pointed out earlier) can
be obtained in a straightforward manner, since they simply express
conservation of particles. If boundary conditions are such that
particles cannot escape from the system, the population of both
species is conserved. This is for example the case when we impose
periodic boundary conditions or also when extremities of the chain are
fixed.

Suppose we fix the total amount $n_a$ and $n_b$ of particles
$(a)$ and $(b)$, this results then on the random walk by the fact that
$n_0+n_3=n_a$ and $n_0+n_1=n_b$ are fixed ($n_i$ is the number of
links $i$). Since $n_0+n_1+n_2+n_3=N$ it then easy to convince oneself
that this is equivalent to fix $n_0-n_2$ and $n_1-n_3$, which is
enough to determine the respective positions of the initial and final
points of the walk.

\begin{figure}[htb]
\begin{center}
\begin{tabular}{cc}
\resizebox*{!}{.4\textwidth}{ \input{gstate2.pstex_t}}&
\hspace{1cm}\resizebox*{!}{.35\textwidth}{ \input{gstate.pstex_t}}\\
(a)&(b)\\[0.2cm]
\includegraphics[width=0.4\textwidth]{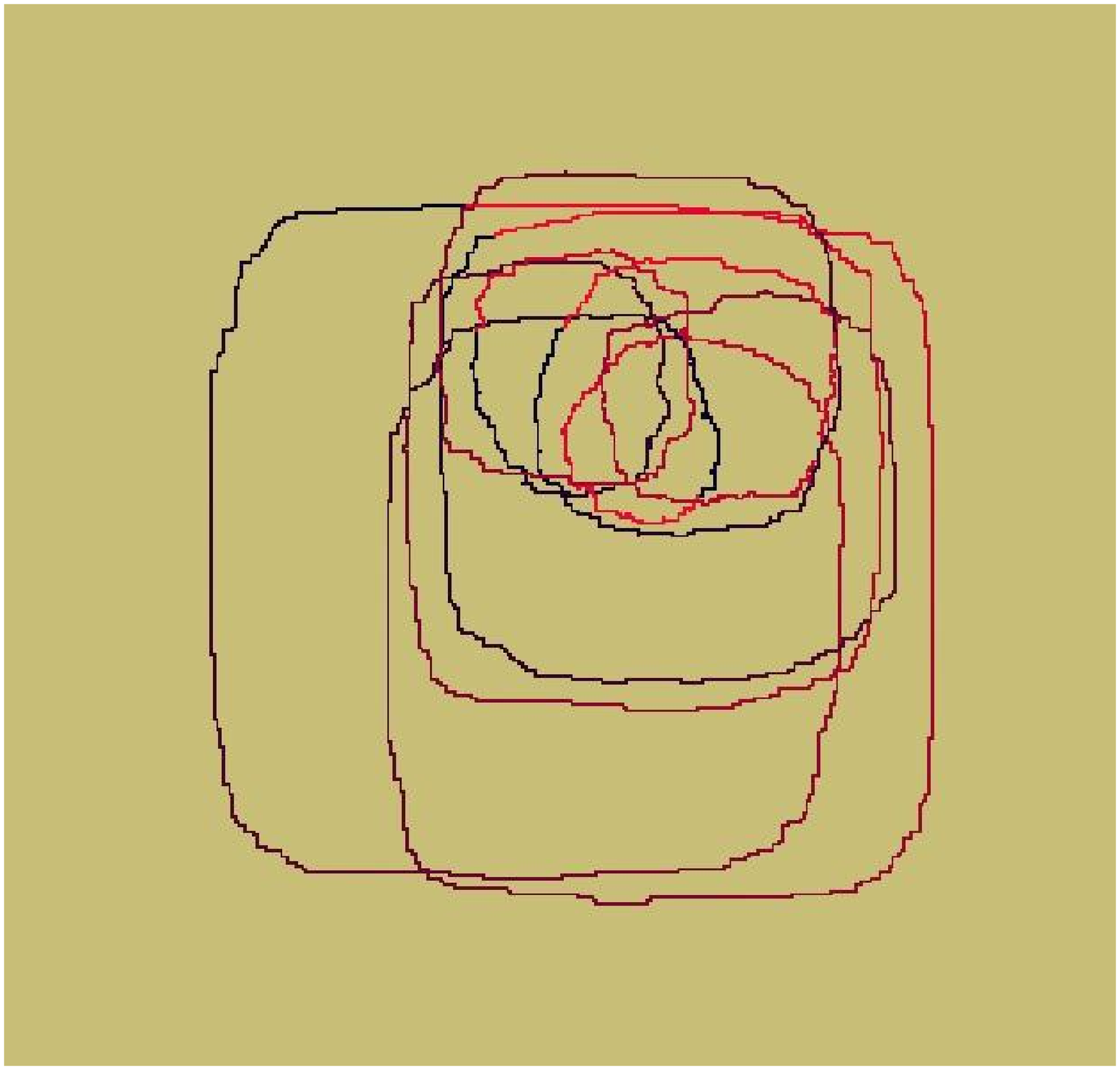}&
\hspace{1cm}\includegraphics[width=0.4\textwidth]{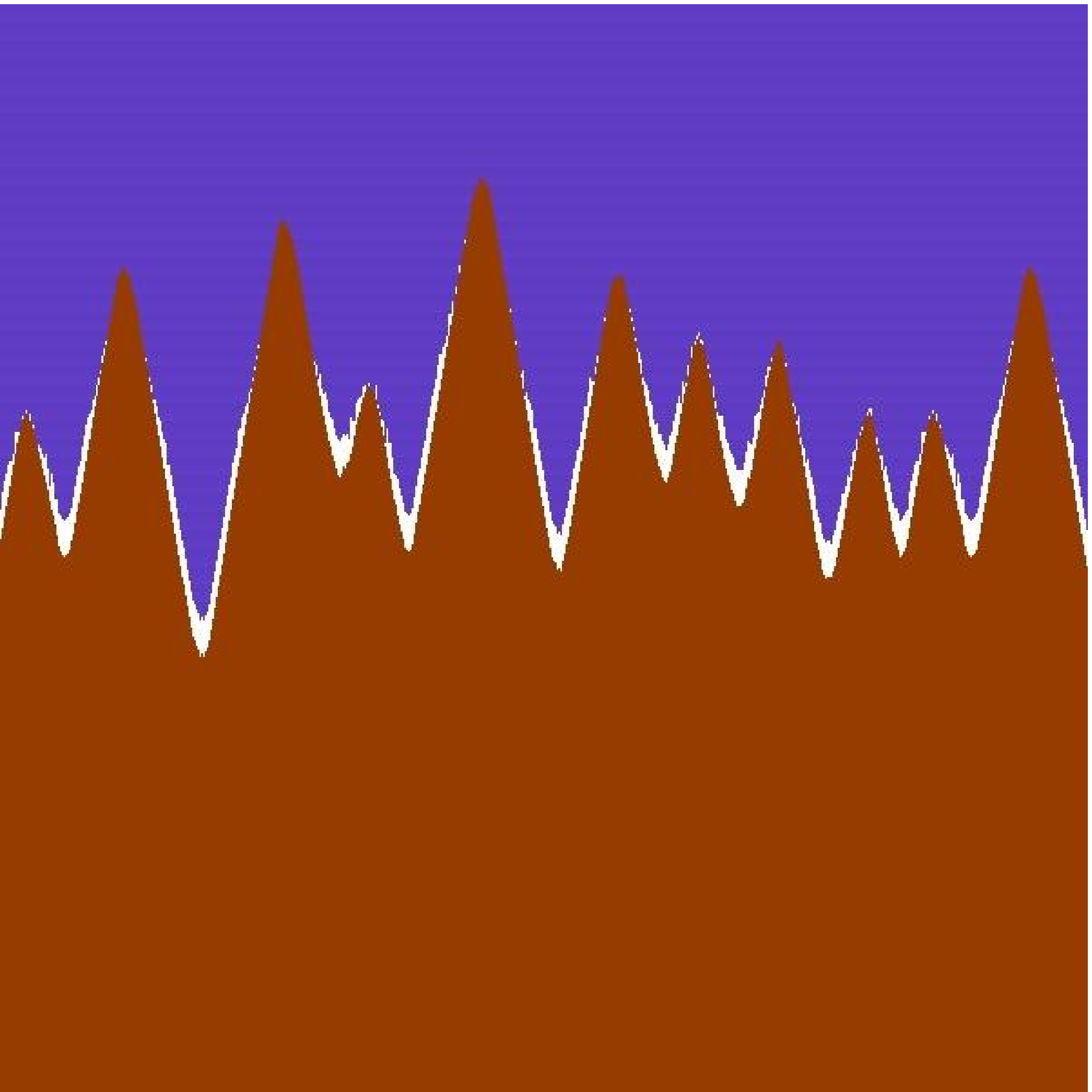}\\
(c)&(d)
\end{tabular}\caption{\label{gstate}(a) Stable configuration for closed chains. (b)
Corresponding representation in terms of exclusion models. (c) Glassy
state with $N=5000$. (d) Corresponding KPZ landscape with the density
of trapped particles represented in white.}
\end{center}
\end{figure}
In addition, except when $\gamma=0$, we get an easy way to determine
the irreducible classes of the system. In fact, for closed systems
with a fixed number of particles, once the population of each species
is fixed, just start from the configuration $(11..1100..00)$ for both
species where all particles have been disposed to the left. Then,
owing to possible consecutive jumps to the right, which for
$\gamma\neq0$ are always authorized, one can reach any arbitrary
configuration. This shows that for closed systems irreducible classes
are indexed by the number of particles in each species, which
corresponds to the separation between extremities of the walk. On the
other hand, for open sample paths with free boundary conditions,
irreducibility holds as long as particles can both enter and leave the
system. An interesting point is the way stable configurations, which
are numerically observed, are represented by means of this exclusion
process formulation. In agreement with the intuition,
Figure~\ref{gstate} depicts a stable situation where particles are
trapped in a well. As this remains true with the complementary
representation (see the lower part of Figure~\ref{gstate}b), the
following iterative scheme could be used to generate the global
invariant measure: let particles (a) evolve assuming the dynamics of
particles (b) is frozen; then, once this \emph{conditional} stationary
regime is reached, switch to particles $(b)$ conditioned by particles
$(a)$, etc. Translated mathematically, we
proposed the following iterative scheme in order to capture the
invariant measure 
\bea 
X^{n+1} &=&
\lim_{t\to\infty}\EE\left(
S^a(t)\mid S^b(t)=Y^n,S^a(0)=X^n\right), \nonumber\\
Y^{n+1} &=& \lim_{t\to\infty}\EE\left(
S^b(t)\mid S^a(t)=X^n,S^b(0)=Y^n\right). \nonumber
\eea
Numerically, the sequences of random variables $X^n$ and
$Y^n$ seemingly converge for
$\eta<50$ to the stationary variables
$S^a(\infty)$ and $S^b(\infty)$.

Up to an abuse of notation, we shall often identify the random process with any of its
sample paths. For instance we shall simply write  
$P(S^a\mid S^b)$ 
[resp. $P(S^b\mid S^a)$] the conditional 
invariant measure of particles (a) [resp. (b)] when the dynamics of 
particles (b)
is frozen. Then the iterative
scheme can be reformulated as
\bea
P_{n+1}(S^a) &=& \sum_{\{S^b\}} P(S^a\mid S^b) Q_n(S^b) \nonumber\\
Q_{n+1}(S^b) &=& \sum_{\{S^a\}} P(S^b\mid S^a) P_n(S^a) \nonumber
\eea
where $P_n$ [resp. $Q_n$] represents the probability measure 
of the $S^a$ [resp. $S^b$] after $n$ steps. If this iteration
converges, the joint probability for the invariant measure 
will be given by
\[ 
P(S^a,S^b) = P(S^a\mid S^b) Q (S^b) = P(S^b\mid S^a) P(S^a)
\]
in case  these two expressions coincide.

\subsection{Conditional equilibrium}

\subsubsection{The case of a stretched walk}\label{special}

Let us have a look to a special case which can be solved exactly. It
will provide some hints about  scaling of the parameters
when $N\to\infty$. 

Consider a random walk with fixed extremities, and consisting only of
links oriented either to the north or to the east, i.e
($\alpha_i\in\{0,1\}, i=1\ldots N$).
\begin{figure}[htb]     
\begin{center}
\resizebox*{!}{4cm}{ \input{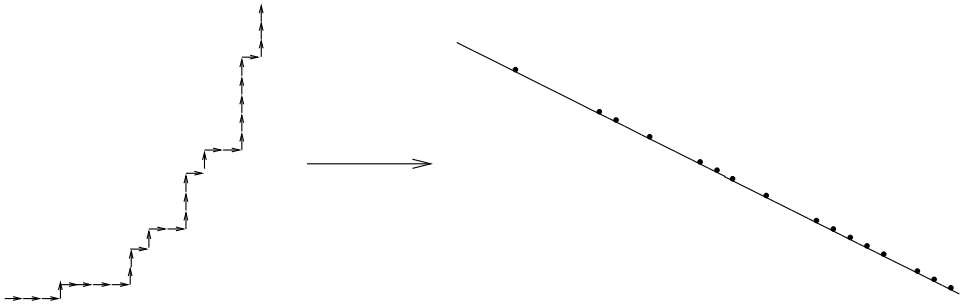}}
\caption{\label{tendu} Stretched random walk and asymmetric exclusion
process.}
\end{center}
\end{figure}
Here  folds do not exist (see figure
(\ref{tendu})), so that $s^b=0$ everywhere and
solely transitions $\lambda^{\pm}$ take place.
It turns out that the invariant measure has the product-form 
\[ P(\alpha_1,\ldots,\alpha_N)=p_{\alpha_1}p_{\alpha_2}...p_{\alpha_N}.\]
Indeed, introducing the occupation rate of particles (a)
$q_i=p_1(i)=1-p_0(i)$, we have the balance equations
\be\label{detbal} \lambda^+(1-q_i)q_{i+1}=\lambda^-q_i(1-q_{i+1}).
\end{equation}
Setting
\[r_i \egaldef\frac{q_i}{1-q_i},
\] 
we obtain a geometric series \be\label{RI}
r_i=r_0\left(\frac{\lambda^-}{\lambda^+}\right)^i =
r_0\exp\biggl[i\log\frac{1-\eta/N}{1+\eta/N}\biggr].
\end{equation}
In (\ref{RI}), the ratio
\[\frac{\eta}{N}=\frac{\lambda^+-\lambda^-}{\lambda^++\lambda^-}= 
\frac{\mu}{\lambda},
\]
already introduced in (\ref{eta}) and (\ref{LM}), gives the
\emph{typical scale} $n_c\simeq\frac{N}{\eta}$, above which we obtain
straight aligned patterns. Here $r_0$ is a normalizing constant, which
permits to fix the expected value $D_0$ of the particle density
\[ \nu = \frac{1}{N}\sum_{i=1}^N \ind{\alpha_i=1}\,,
\] 
 that is 
\[ D_0 = \mathbb{E}(\nu)=\frac{1}{N}\sum_{i=1}^N
q_i=\frac{1}{N}\sum_{i=1}^N \frac{r_i}{1+r_i} .
\]
Letting $N\to\infty$, we analyze the limiting process under the
\emph{scaling}, 
\begin{equation} \label{SCALE}
\eta \egaldef \lim_{N\to\infty} \frac{N\mu}{\lambda},
\end{equation}
where $\eta$, up to some abuse in the
notation, is  now a parameter independent of $N$. This is tantamount to assume
that $\mu$ is implicitly a function of $N$.

Fixing $x= i/N$ and taking the expansion with respect to $\eta/N$ in
(\ref{RI}), we get the limit equation
\be\label{rdx} r(x)=r_0\exp(-2\eta x),
\end{equation}
which implies in turn
\[ D_0 = \int_0^1 dx\left(1-\frac{1}{1+r_0\exp(-2\eta x)}\right)
=\frac{1}{2\eta}\log\frac{1+r_0}{1+r_0\exp(-2\eta)}.
 \]
Consequently, 
\[r_0=\frac{\sinh(\eta D_0)}{\sinh\eta(1-D_0)}\ e^\eta.\] 
The integral of the particle density, taken as a function of $x$, is
shown in Figure~\ref{tendu} and is given by
\[ h(x) = \int_0^x du\left(1-\frac{1}{1+r_0\exp(-2\eta u)}\right) 
= \frac{1}{\eta}\log\left[\frac{1+r_0}{1+r_0\exp(-2\eta x)}\right].
\]
This asymmetric exclusion model can be solved under more general
conditions, in particular with open boundaries, using matrix methods
\cite{DERRIDA}. The above simple example confirms the observed fact
that the correct scaling parameter is indeed $\eta$, and also somehow
explains why the chain remains Brownian when $\eta\lesssim1$.

\subsubsection{The general case}
Once the sequence $\{s^b\}$ is given, particles (a) form a
unidimensional exclusion process in an inhomogeneous environment. Its
transition rates to the right or to the left at position $i$ have been
given in (\ref{taux}). 

Our basic claim is that the reasoning of section \ref{special} still
holds for a closed system where the number of particles is kept
constant. Exactly this means that, as long as there is no
\emph{current}  in the system, conditional detailed
balance equations of type (\ref{detbal}) are still valid at steady
state, just replacing $\lambda^\pm$ by $\lambda_a^\pm(i)$. 

We present no proof of this fact (which could likely be obtained by
coupling arguments), and write brutally the equilibrium equations
\[ 
\lambda_a^+(i)(1-q_i^a)q_{i+1}^a=\lambda_a^-(i)q_i^a(1-q_{i+1}^a),\qquad
i=1\ldots N,
\]
where sites $N+1$ and $1$ are identified for periodic boundary
conditions, and $q_i^a$ [resp. $q_i^b$] is  the  random variable
equal to  the
conditional probability of having one particle of type (a) [resp. (b)]
in position $i$ given the sequence $\{s^b\}$ [resp. $\{s^a\}$], that is
\begin{equation} \label{COND}
q_i^a= \EE [s_i^a \mid S^b], \quad q_i^b = \EE [s_i^b \mid S^a].
\end{equation}

Setting
\[
r_i^a\egaldef\frac{q_i^a}{1-q_i^a},
\]
we obtain 
\be\label{differ} 
\log[r_{i+1}^a]-
\log[r_i^a]=\log\biggl(\frac{\lambda_a^-(i)}{\lambda_a^+(i)}\biggr).
\end{equation}
By using (\ref{taux}),  an easy algebra based on the boolean
character of the $s_i^b$ yields
\begin{equation}\label{differ2}
\log\frac{\lambda_a^-(i)}{\lambda_a^+(i)} = 
(1-2s_i^b)\log\frac{\lambda-\mu}{\lambda+\mu} +
(s_{i+1}^b- s_i^b)\log \frac{(\gamma-\delta)(\lambda+\mu)}{(\gamma+\delta)
(\lambda-\mu)},
\end{equation}
with a similar equation for type $b$ particles. In addition we observe
that the constraints
\[r_{N+1}^a=r_1^a, \quad r_{N+1}^b=r_1^b,
\]
 for periodic boundary
conditions, will be automatically fulfilled as long as the system of
particles (b) is globally neutral (i.e. particles and holes have the
same cardinality), in which case
\[\sum_{i=1}^N(1-2s_i^b) = 0.\]

\subsection{Closed curves: weak convergence for large $N$}\label{SECW}
We will now combine the stationary product forms obtained for each
particle species, according to the iterative scheme proposed at the
end of section \ref{Mapping}. Throughout this section, the dependence
on $N$ of the random variables $q_k^a, q_k^b$, given by (\ref{COND}),
is kept implicit for the sake of shortness in the notation.

Introduce  $\{w_i^a, i\ge 0\}$ and $\{w_i^b, i\ge 0\}$, two families of
 independent and identically distributed Bernoulli random variables
 with parameter $1/2$ and values $\{1,-1\}$. Without further comment,
 we assume that $(S^a,S^b),\{w_i^a, i\ge 0\}, \{w_i^b, i\ge 0\}$ are
 defined on the same probability space.

\begin{lem}\label{WEAK}
 Let $\alpha_k, k\ge 1$, denote a sequence of complex numbers
satisfying the condition $\sup_k |\alpha_k|<\infty$.

There exists a probability space such that
\begin{equation}\label{COUPLE}
\begin{cases}
\DD\frac{1}{N}\sum_{k=1}^{N}\alpha_ks_k^a =
\frac{1}{N}\sum_{k=1}^{N}\alpha_k (q_k^a +\sigma_k^aw_k^a) + O(N^{-2})
\quad a.s. ,\\[0.5cm]
 \DD\frac{1}{N}\sum_{k=1}^{N}\alpha_ks_k^b =
\frac{1}{N}\sum_{k=1}^{N}\alpha_k (q_k^b +\sigma_k^bw_k^b) +
O(N^{-2})\quad a.s.,
\end{cases}
\end{equation}
 where
\begin{equation}\label{sigma} 
\sigma_k^a = \sqrt{q_k^a(1-q_k^a)},\quad 
\sigma_k^b = \sqrt{q_k^b(1-q_k^b)}, \quad\forall k\ge 1.
\end{equation}
\end{lem}

\begin{proof}
The starting point is the straightforward decomposition of the
probability measure $P(S^a,S^b)$ as
\[ P(S^a,S^b)=P(S^a \mid S^b)P(S^b)=P(S^b\mid S^a)P(S^a). 
\]
Considering the invariant measure of the process $(S^a(t),S^b(t))$, we
claim  the conditional probabilities $P(S^a\,|\,S^b)$ and $P(S^b
\mid S^a)$ coincide with the equilibrium probabilities obtained in the
last section \emph{for each particle species}. In the present
situation, this is tantamount to writing the equation
\[ P(S^a \mid S^b) = \prod_{i=1}^N \bigl(s_i^a q_i^a +
\bar s_i^a(1-q_i^a)\bigr), \] where the $q_i^a$\,'s depend implicitly
of $S^b$. To analyze more precisely the coupling between the two
families, we introduce the Laplace transforms
\[
\varphi_k^a(\alpha) \egaldef \EE \biggl[\exp
\biggl\{\frac{1}{N}\sum_{k=1}^N\alpha_k s_k^a\biggr\}\biggr] =
\EE\biggl[\prod_{k=1}^N\bigl[1+q_k^a(e^{\frac{\alpha_k}{N}}-1)\bigr]\biggr].
\]
Then
\[
\varphi_k^a(\alpha)= \EE\biggl[\exp \biggl\{\frac{1}{N}\sum_{k=1}^N\alpha_k
q_k^a + \frac{1}{2N^2}\sum_{k=1}^N \alpha_k^2
q_k^a(1-q_k^a)+O(\frac{1}{N^2})\biggr\}\biggr].
\] 
On the other hand, starting from the equality
\[
\EE\biggl[\exp\biggl\{\frac{1}{N}\sum_{k=1}^n \alpha_k(
q_i^a+\sigma_i^aw_i^a)\biggr\}\biggr] =
\EE\biggl[\exp\biggl\{\frac{1}{N}\sum_{k=1}^N \alpha_kq_k^a +
\sum_{k=1}^N\log\cosh\frac{\sigma_k^a\alpha_k}{N}\biggr\}\biggr],
\]
with regard to the (a) species, we observe that the value of
$\sigma_k^a$ given in (\ref{sigma}) gives the matching
\begin{equation}\label{EQUIV}
\EE\biggl[\exp\biggl\{\frac{1}{N}\sum_{k=1}^N\alpha_k
s_k^a\biggr\}\biggr] =
\biggl[\EE\exp\biggl\{\frac{1}{N}\sum_{k=1}^N\alpha_k
(q_k^a+\sigma_k^a w_k^a)+O(\frac{1}{N^2})\biggr\}\biggr].
\end{equation}

Since all random variables at stake are uniformly bounded, equation
 (\ref{EQUIV}) yields at once (\ref{COUPLE}), but only in
 distribution. To conclude the proof of the lemma, we make use of
 transfer and coupling theorems due to Skorohod and Dudley (see
 \cite{KAL}, theorem 4.30 and corollaries 6.11, 6.12), which allow to
 switch from equalities in distribution to almost sure properties,
 since on the original probability space the right member of system
 (\ref{COUPLE}) is a measurable mapping of the left one.
\end{proof}

\subsubsection{Fundamental scaling, thermodynamic limit and fluctuations}
For any $i$, $1\le i\le N$, we put ad libitum $x=i/N,0\le x\le 1$.
\begin{prop}\label{LOTKA} Under
the \emph{fundamental scaling}
\[ 
\begin{cases}
\DD \frac{\mu}{\lambda} = \frac{\eta}{N}+o\Bigl(\frac{1}{N}\Bigr),\\[0.4cm]
\DD \frac{\delta}{\gamma}=O\Bigl(\frac{1}{N}\Bigr),
\end{cases}
\]
the weak limits  
\begin{equation} \label{RHO}
\rho^a(x) = \lim_{N\to\infty} q_{xN}^a \quad \mathrm{and} \quad 
\rho^b(x) = \lim_{N\to\infty} q_{xN}^b
\end{equation}
exist and satisfy the autonomous system of deterministic nonlinear
differential equations
\begin{equation}\label{deter}
\begin{cases}
\DD\frac{\partial\rho^a(x)}{\partial x} =
4\eta\rho^a(x)(1-\rho^a(x))\Bigl(\rho^b(x)-\frac{1}{2}\Bigr), \\[0.5cm]
\DD\frac{\partial\rho^b(x)}{\partial x} =
-4\eta\rho^b(x)(1-\rho^b(x))\Bigl(\rho^a(x)-\frac{1}{2}\Bigr). 
\end{cases}
\end{equation}
In addition, the assumed closure of the original random walk imposes
 the relations
\begin{equation} \label{PERIOD}
\begin{cases} 
\int_0^1 \rho^a(x)dx = \int_0^1 \rho^b(x)dx =\frac{1}{2}, \\[0.1cm]
\rho^a(x+1) = \rho^a(x),\\ 
\rho^b(x+1) = \rho^b(x).   
\end{cases}
\end{equation}
\end{prop}

\begin{proof}
Taking the expansion with respect to $N$ in equations
(\ref{differ}) and (\ref{differ2}), and using lemma
\ref{WEAK}, we get after some algebra

\begin{equation}\label{scaleq} 
\begin{cases}
\DD \log \frac{r_k^a}{r_1^a}=
\frac{2\eta}{N}\sum_{j=1}^{k-1}(2q_j^b+2\sigma_j^b w_j^b -1) +
O\Bigl(\frac{1}{N}\Bigr) \quad a.s. , \\[0.4cm] 
\DD \log \frac{r_k^b}{r_1^b} =
-\frac{2\eta}{N}\sum_{j=1}^{k-1}(2q_j^a+2\sigma_j^a w_j^a -1) +
O\Bigl(\frac{1}{N}\Bigr)\quad a.s.
\end{cases}
\end{equation}

Since $N$ is a parameter and $x$ rather stands for a variable, it is
be convenient to introduce the following functions of $x$
\[ 
q_N^a(x) \egaldef q_{[xN]}^a, \qquad r_N^a(x)
\egaldef\frac{q_N^a(x)}{1-q_N^a(x)},\qquad
\sigma_N^a(x)\egaldef\sqrt{q_N^a(x)(1-q_N^a(x))},
\]
and similarly for the (b) species. 

To omit some tedious technicalities, we will only sketch the remaining 
lines of the proof.

First, it is not difficult to see that by  restricting  the expansion
 in (\ref{EQUIV}) up to terms of order $N^{-1}$, and using the
 definition of $r_k^a, r_k^b$, we come to the simplified system 
\begin{equation}\label{QKA} 
\begin{cases}
 q_k^a = \DD\frac{r_1^a\exp
\Bigl[\frac{2\eta}{N}\DD\sum_{j=1}^{k-1}(2q_j^b -1)\Bigl]}{1 + r_1^a\exp
\Bigl[\frac{2\eta}{N}\DD\sum_{j=1}^{k-1}(2q_j^b -1)\Bigr]} +
O\Bigl(\frac{1}{N}\Bigr) \quad a.s. , \\[0.6cm] 
 q_k^b =\DD
\frac{r_1^b\exp \Bigl[-\frac{2\eta}{N}\DD\sum_{j=1}^{k-1}(2q_j^a
-1)\Bigl] }{1 + r_1^b\exp \Bigl[-\frac{2\eta}{N}\DD\sum_{j=1}^{k-1}(2q_j^a
-1)\Bigr]} + O\Bigl(\frac{1}{N}\Bigr) \quad a.s.
\end{cases}
\end{equation}

In  a second step, it can be shown from (\ref{QKA}), as in a purely
 deterministic context, that the quantities $q_N^a(x),q_N^b(x)$, form
 Cauchy sequences, hence converging, for all $0\le x\le 1$. To see
 that the deterministic limits (\ref{RHO}) exist and satisfy
 (\ref{deter}) is straightforward by approximating discrete sums by
 Riemann's integrals. This yields the differential system
\[
\begin{cases}
\DD\frac{\partial}{\partial
x}\bigl[\log\frac{\rho^a(x)}{1-\rho^a(x)}\bigr] = 2\eta
(2\rho_b(x)-1),\\[0.3cm]
\DD\frac{\partial}{\partial
x}\bigl[\log\frac{\rho^b(x)}{1-\rho^b(x)}\bigr] = -2\eta
(2\rho_a(x)-1),
\end{cases}
\]
which has similarities with the famous Lotka-Volterra equations, where
$x$ plays here the role of the time. It is worth noting that
(\ref{rdx}) is immediately revisited, taking merely
$\rho^b(x)\equiv 0$ (i.e. the density of particles (b) is kept
constant).
\end{proof}

It is also tempting to get some insight into fluctuations around the
above deterministic limit. This will be achieved by establishing the
forthcoming
central limit theorem. 
\begin{prop}\label{CLT}
Under the \emph{fundamental scaling}, the weak limits 
\begin{equation}\label{FLUC}
g^a(x)=\lim_{N\to\infty}
\sqrt{N}\,\frac{q_{N}^a(x)-\rho^a(x)}{\sigma_a^2(x)},\quad
g^b(x)=\lim_{N\to\infty}
\sqrt{N}\,\frac{q_{N}^b(x)-\rho^b(x)}{\sigma_b^2(x)},
\end{equation}
exist and satisfy the system of stochastic differential
equations
\begin{equation}\label{stoch2}
\begin{cases}
dg^a(x)=4\eta\Bigl[\sigma_a^2(x)g^b(x)dx\ +\
\sigma_b(x)dW^b(x)\Bigr],\\[0.2cm]
dg^b(x)=-4\eta\Bigl[\sigma_b^2(x)g^a(x)dx\ +\
\sigma_a(x)dW^a(x)\Bigr],
\end{cases}
\end{equation}
where 
\[\sigma_a(x)\egaldef\sqrt{\rho^a(x)(1-\rho^a(x))}, \quad 
\sigma_b(x)\egaldef\sqrt{\rho^b(x)(1-\rho^b(x))}.
\]
\end{prop}

\begin{proof}
Letting
\[
X^a_N(x) \egaldef \frac{1}{\sqrt N}\sum_{k=1}^{[xN]} w_k^a, \qquad
X^b_N(x) \egaldef \frac{1}{\sqrt N}\sum_{k=1}^{[xN]} w_k^b,
\]
we consider the two white noise processes
\[
W^a(x) \egaldef \lim_{N\to\infty} X^a_N(x),\qquad 
W^b(x) \egaldef \lim_{N\to\infty} X^b_N(x).
\]
We call on a strong approximation theorem, which is a refinement of
the invariance principle (see \cite{KOMLOS}: it shows how to construct
$W^a(x)$ and $W^b(x)$ on the same probability space as $X^a_N$ and
$X^b_N$) in such a way that
\[
\sup_{0\le x\le 1} |X_N(x) - W(x)| =  O\Bigl(\frac{\log N}{\sqrt {N}}\Bigr).
\]

Arguing as in the derivation of proposition \ref{LOTKA}, we can write
\begin{equation}\label{stoch}
\begin{cases}
\DD\log \frac{r_N^a(x)}{r_N^a(0)} =  2\eta\int_0^x du(2 q_N^b(x)-1) +
\frac{4\eta}{\sqrt N}\int_0^x \sigma_N^b(u) dW^b(u) +
O\Bigl(\frac{\log N}{N}\Bigr),\\[0.5cm] \DD
\DD\log \frac{r_N^b(x)}{r_N^b(0)} =  -2\eta\int_0^x du(2 q_N^a(x)-1) -
\frac{4\eta}{\sqrt N}\int_0^x \sigma_N^a(u) dW^a(u) +
O\Bigl(\frac{\log N}{N}\Bigr),
\end{cases}
\end{equation}
 where the stochastic integrals are taken in the ItÙ sense (see
 \cite{OKSENDAL}). Defining  $g_N^a(x)$ and $g_N^b(x)$ by the equations
\begin{eqnarray*}
q_N^a(x)&=&\rho^a(x)+\sigma_a^2(x)g_N^a(x),\\
q_N^b(x)&=&\rho^b(x)+\sigma_b^2(x)g_N^b(x),
\end{eqnarray*}
then putting these expressions into (\ref{stoch}) and differentiating
 (\ref{stoch}) (details are omitted), we are lead to (\ref{stoch}.
The proof of proposition \ref{CLT} is completed.
\end{proof}

\paragraph{Remark} Let us comment on   the scaling of $\delta$.  
Contrary to the scaling of $\mu$, which is naturally dictated by
homogeneity (the sums in (\ref{scaleq}) remain meaningful after
dividing by $N$, when $N\to\infty$), it is a purely dynamical
consideration which dictates the scaling of $\delta$. Indeed, $\mu$
and $\delta$ are associated to time constants $\tau_{\mu}$ and
$\tau_{\delta}$. The quantity $\tau_{\mu}$ represents the typical unit
of time for a free particle (a) or (b) to drift along the system over
a finite distance, whereas $\tau_{\delta}$ is a time-scale for
rotations of vertical or horizontal fold $M2$ of the chain,
remembering that $\delta$ stands for the \emph{detuning} between
$\gamma^+$ and $\gamma^-$ defined in section \ref{Introduction}.
Therefore either these time-scale are coherent and $\delta$ is
rescaled, otherwise rotational motions of $M2$ occur at a shorter time
scale and a different analysis has to be conducted, since motifs $M2$
(which correlate hole-particle pairs of species (a) and (b)) reach
their equilibrium distribution before particles have enough time to
move along the system. Numerically we could not perceive any specific
effect related to $\delta$, so that we restricted ourselves to the
above \emph{fundamental scaling}.

\subsubsection{Second order phase transition }
Here we focus on on the deterministic (\ref{deter}) part of the
equations. Using the notation
\[ \nu_a(x) \egaldef 2\rho^a(x)-1,\qquad \nu_b(x) \egaldef 2\rho^b(x)-1, \]
apart from the trivial solution $\nu_a(x)=\nu_b(x)=0$,
we obtain from  (\ref{deter}) 
\[ \frac{\nu_a(x)}{1-\nu_a^2(x)}\frac{\partial \nu_a(x)}{\partial x}=
-\frac{\nu_b(x)}{1-\nu_b^2(x)}\frac{\partial \nu_b(x)}{\partial x}, \]
or, after integration, 
\[ 1-\nu_a^2(x) = \frac{C}{1-\nu_b^2(x)}, \qquad 0<C<1, \]
since $|\nu_a|<1$ and  $|\nu_b|<1$.
Plugging  the last relation into  (\ref{deter}) leads to
\be\label{diff1}
\biggl(\frac{\partial \nu_a(x)}{\partial x}\biggr)^2 
= \eta^2[1-\nu_a^2(x)][1-C-\nu_a^2(x)],
\end{equation}
the solution of which is the standard Jacobi elliptic function
\[
\nu_a(x) = \frac{1}{\sqrt{1-C}}\, \mathrm{sn} (\eta x,\sqrt{1-C}). 
\]
Finding the constant $C$ is equivalent to compute the fundamental period
of these functions. Hence, denoting by $X(C)$ the period of $\nu_a(x)$, we
have solve
\begin{equation}\label{PER}
X(C)=1.
\end{equation}
 From (\ref{diff1}) we see that $\nu_a(x)$ is
bounded by $-\sqrt{1-C}$ and $\sqrt{1-C}$. Taking into account the
 constraint ($\int_0^1\nu_a(x)dx =0$), $X(C)$ is exactly given by
\[
X(C)= \frac{1}{\eta} F(\frac{\pi}{2},\sqrt{1-C}) =
\frac{4}{\eta}\int_0^1\frac{d\nu}{\sqrt{[1-\nu^2][1-(1-C)\nu^2]}}, \]
where $F$ is the standard elliptic integral of first kind.

$X(C)$ is a decreasing function of $C$ on $]0,1]$, reaching its
minimum for $C=1$, so that
\[
X(C)\ge X(1) = \frac{2\pi}{\eta}. \]
Thus appears is a \emph{critical value} for $\eta$, namely
\[ \eta_c = 2\pi.
 \] 

When $\eta_c <2\pi$, (\ref{PER}) cannot be fulfilled and we are left
with the trivial solution.

When $\eta_c \ge 2\pi$, it is straightforward to compute the
arithmetic area $S_1(\eta)$ of the first winding sector, since it is
indeed the only non-vanishing sector. Setting
\[
\begin{cases}
\DD h_x(u) = \frac{1}{2}\int_0^u dv[\nu_a(v)+\nu_b(v)] dv,\\[0.3cm]
\DD h_y(u) = \frac{1}{2}\int_0^u dv[\nu_a(v)-\nu_b(v)] dv,
\end{cases}
\]
$S_1(\eta)$ is simply the area enclosed by the curve $(h_x(u),h_y(u))\
u\in[0,1]$, which is given by
\[
S_1(\eta) = \frac{1}{2}\int_0^1 du \Bigl[h_x\frac{\partial h_y}{\partial u}
- h_y\frac{\partial h_x}{\partial u} \Bigr],
\]
or, after some algebra,
\[
S_1(\eta) = \frac{1}{2\eta^2}\int_0^{\sqrt{1-C}}
\frac{\nu}{\sqrt{(1-\nu^2)(1-C -\nu^2)}}
\log\Bigl[\frac{1+\nu}{1-\nu}\Bigr] d\nu ,
\]
keeping in mind that $C$ is also a function of $\eta$. The corresponding
curve displayed in figure \ref{etac}.b matches pretty nicely all
numerical observations, in particular with regard to the critical value
$\eta_c$.

\subsubsection{More about fluctuations for $\eta<\eta_c$}
 When  $\eta$ is under the threshold  $\eta_c$,   
the  deterministic part becomes  trivial,
and we are left with fluctuations. This corresponds basically to 
the observations shown in Figures \ref{phases}.a and \ref{phases}.b.
Inserting
\[\rho^a(x)=\rho^b(x)=\frac{1}{2},\]
in (\ref{stoch2}), and setting 
\begin{equation}\label{BRO}
g(x) = g^a(x) + ig^b(x), \qquad W(x)=W^a(x)+iW^b(x),
\end{equation}
we get solutions of the form
\[
g(x) = -2i\eta\int_0^x e^{i\eta(u-x)}dW(u), 
\] 
which corresponds to
\begin{eqnarray*} 
q^a(x)=\frac{1}{2}+\frac{\eta}{\sqrt N}\Bigl[
\int_0^x\bigl(\sin(\eta(u-x)) dW^a(u)+\cos(\eta (u-x)) dW^b(u)\bigr)\
+\ dW^a(x)\Bigr] + o\Bigl(\frac{1}{\sqrt N}\Bigr), \\[0.2cm] 
q^b(x)=\frac{1}{2}+\frac{\eta}{\sqrt
N} \Bigl[\int_0^x\bigl(\cos(\eta (u-x)) dW^a(u)-\sin(\eta (u-x))
dW^b(u)\bigr)\ +\ dW^b(x)\Bigr] + o\Bigl(\frac{1}{\sqrt N}\Bigr).
\end{eqnarray*}
 Hence we have derived an \emph{equivalent} process, which up to order
$o(\frac{1}{\sqrt N})$, describes the curves numerically observed.
Letting
\[
\begin{cases}
\DD h_N^a(x) \egaldef \frac{1}{N}\sum_{j=1}^{[xN]} (2s_j^a-1), \\[0.5cm]
\DD h_N^b(x) \egaldef \frac{1}{N}\sum_{j=1}^{[xN]} (2s_j^b-1),
\end{cases}
\]
the so-called equivalence says precisely
\[
dh_N^a(x)= h_N^a\bigl(x+\frac{1}{N}\bigr)-h_N^a\bigl(x\bigr)=
2q_N^a(x)+\frac{2}{\sqrt N}\sigma_N^a(x)dW^a(x)-1+
o\Bigl(\frac{1}{\sqrt N}\Bigr)
\]
and the same holds for $dh_N^b(x)$. Introducing the complex function
\[
h_N(x) = h_N^a(x) + ih_N^b(x),\]
we have 
\[ dh_N(x)= -\frac{2i\eta}{\sqrt N}\Bigl[\int_0^x e^{i\eta(u-x)}dW(u)\
+ \ dW(x)\Bigr] +  o\Bigl(\frac{1}{\sqrt N}\Bigr),
\] 
which, after integrating by parts, yields
\[
h_N(x)= \frac{1}{\sqrt N}\Bigl[\int_0^x 2e^{i\eta(u-x)} dW(u) -
W(x)\Bigr] + o\Bigl(\frac{1}{\sqrt N}\Bigr).
\] 
At this point it is possible to reconstruct the curves observed
numerically, remembering that the discrete displacements $(dx_i,dy_i)$
in the plane are expressed in terms of $s_i^a$ and $s_i^b$ as
\[
\begin{cases}
dx_i = 1-s_i^a-s_i^b, \\
dy_i = s_i^a-s_i^b.
\end{cases}
\]
According to (\ref{BRO}), we define 
\[
h(x)=\lim_{N\to\infty}\frac{1}{\sqrt N} \sum_{k=0}^{[xN]} (dx_k + i
dy_k)  \quad \mathrm{and} \quad Z(x) = \frac{i-1}{2}W(x).
\]
Then we have 
\[
h(x)=\int_0^x 2 \exp [i\eta(u-x)]
dZ(u) - Z(x).
\]
This above equation
accounts for the windings of the Brownian curve observed in figure
\ref{phases}.b. When $\eta\to 0$, $h(x)$ coincides
with the standard Brownian motion $Z$.

\subsection{Burgers equations in the fluid limit}

In this section with give a formal derivation of dynamical equations
 describing the fluid limit, which includes the steady state solutions
 obtained in \ref{SECW}, without  pretending to a rigourous
 presentation.
 
We start from the heuristic assumption that the conditional
 independence of the $s_i^a$ [resp. $s_i^b$], which is realized at
 time $t=0$ and at steady state, remains valid for all fixed time $t$,
 up to order $O(N^{-1})$. Exact proofs of this fact could be provided
 by adapting (up to sharp technicalities) some lines of argument
 proposed e.g. in \cite{LIG,DUR}), together with a mean field type
 approach for the convergence of the semi-groups of the underlying
 Markov processes indexed by $N$.

 Considering the stochastic variable expressing the current of particles
(a) between sites $i$ and $i+1$ at time $t$
\begin{eqnarray*}
 \varphi_i^a(t) &=&
\lambda_a^+(i,t)s_i^a(t)\bar s_{i+1}^a(t) -
\lambda_a^-(i,t)s_{i+1}^a(t)\bar s_i^a(t),\\
\varphi_i^b(t) &=&
\lambda_b^+(i,t)s_i^b(t)\bar s_{i+1}^b(t) -
\lambda_b^-(i,t)s_{i+1}^b(t)\bar s_i^b(t),
\end{eqnarray*}
we define the conditional expectation 
\[
J_i^a(t)= \EE[\varphi_i^a(t) \mid S^b], \quad J_i^b(t) =
\EE[\varphi_i^b(t) \mid S^a].
\]
On account of the particle conservation principle, we have locally
\begin{equation} \label{COND3}
\frac{\partial}{\partial t}\EE[s_i^a(t)]+\EE\bigl[\varphi_i^a(t) -
\varphi_{i-1}^a(t)\bigr] = 0.
\end{equation}
Then introducing the time dependent expectations  
$q_k^a(t)$ as functions 
of the sample path $S^b$ 
\begin{equation} \label{COND2}
q_k^a(t)= \EE[s_k^a(t) \mid S^b], \quad q_k^b(t) = \EE[s_k^b(t) \mid
S^a],
\end{equation}
 we can write by (\ref{COND3})
\begin{equation}\label{DEQ}
\frac{1}{N}\sum_{k=1}^N \alpha_k\Bigl(q_k^a(t)-q_k^a(0)+
\int_0^t d\tau \bigl(J_k^a(t) - J_{k-1}^a(t)\bigr)\Bigr) = 0,
\end{equation}
with again the condition $\sup_k |\alpha_k|<\infty$. Expressing
the \emph{almost} conditional independence of the $s_k^a$'s,  we
have
 \bea
\frac{1}{N}\sum_{k=1}^N \alpha_k\EE_t
\Bigl[\lambda_a^+(k,t)q_k^a(t)(1-q_{k+1}^a(t))-
\lambda_a^-(k,t)q_{k+1}^a(t)(1-q_k^a(t))\Bigr] =&&\nonumber\\
\frac{1}{N}\sum_{k=1}^N \alpha_k J_k^a(t) + o\Bigl(\frac{1}{N}\Bigr) .\nonumber&&
\eea
 To be consistent with the procedure developed for the stationary
regime case, we substitute to the $S^b$ the so-called equivalent set
$\{q_k^b+\sigma_k^b w_k^b\}$ into the expressions of the rates given by
\bea
\log \lambda_a^\pm(i) &&=\  \log \lambda +
2(s_i^b+s_{i+1}^b-2s_i^bs_{i+1}^b) \log\frac{\gamma}{\lambda}\nonumber\\
&&\pm\ \frac{\eta}{N}(1-s_i^b-s_{i+1}^b)
\pm\frac{\delta}{\gamma}(s_i^b-s_{i+1}^b) + o\Bigl(\frac{1}{N}\Bigr),\nonumber
\eea
 according to the \emph{fundamental scaling}. This insures that
$\lambda_a^+$ and $\lambda_a^-$ remain perfectly correlated. As for the
 the fluid limit, the procedure amounts simply to replace
$s_i^b$ by $q_i^b$ in $\lambda_a^\pm(i)$, and to approximate all
discrete sums in (\ref{DEQ}) by Riemann's integrals, for arbitrary
$\alpha(x)$. This yields the continuity equation
\begin{equation}\label{TRANS}
\frac{\partial\rho^a(x,t)}{\partial t} = - \frac{\partial
J^a(x,t)}{\partial x},
\end{equation}
where $\rho_a(x,t)$ and $J^a(x,t)$ 
are the  deterministic continuous counterparts
of $q_k^a(t)$ and $J_k^a(t)$.
The current is now given by
\begin{equation}\label{current}
J^a(x,t) = D \biggl[2\eta\rho^a(1-\rho^a)(1-2\rho^b)
-\frac{\partial\rho^a(x,t)}{\partial x}\biggr]
\exp\Bigl(2\rho^b(1-\rho^b)\log\frac{\gamma}{\lambda} \Bigr),
\end{equation}
where we have introduced the diffusion constant
\[
D \egaldef \lim_{N\to\infty} \frac{\lambda}{N^2},
\]
$\lambda$ being implicitly taken as a function of $N$. This
scaling is confirmed numerically (see figure \ref{etac}.c). Actually we observe
that the parameter $\gamma$ controls the dynamics of the system
through the definition of an effective diffusion constant. For
particles (a), we have
\[
\DD D^a(x,t) \egaldef D
\exp\Bigl[2\rho^b(x,t)(1-\rho^b(x,t))\log\frac{\gamma}{\lambda}
\Bigr].
\]
with the corresponding relation for particles (b)). In the particular
 case $\gamma\to 0$, this constant vanishes, except at loci where the
 density of particles (b) has no fluctuations, that is
 $\sigma^2=\rho^b(1-\rho^b)=0$), in which case $D^a(x,t)=D$, as to be
 expected from the analysis of the stretched walk. When
 $\gamma=\lambda$, we obtain a dynamical system of
 deterministic equations 
\begin{eqnarray*} \frac{\partial\rho^a(x,t)}{\partial
 t}&=&D\frac{\partial^2\rho^a(x,t)}{\partial x^2}
 -2D\eta\frac{\partial}{\partial
 x}\bigl[\rho^a(1-\rho^a)(1-2\rho^b)\bigr](x,t), \\[0.2cm]
 \frac{\partial\rho^b(x,t)}{\partial
 t}&=&D\frac{\partial^2\rho^b(x,t)}{\partial x^2}
 +2D\eta\frac{\partial}{\partial
 x}\bigl[\rho^b(1-\rho^b)(1-2\rho^a)\bigr](x,t). 
\end{eqnarray*}
 These equations belong to Burgers' class. When taking one of the two
 density species (say (a)) to be a constant $\rho_a=0$ or $\rho_a=1$,
 the density of particles (b) is  driven by an ordinary Burgers equation,
 describing  the evolution of a stretched walk. For an arbitrary
 $\gamma$, the steady state solution of (\ref{TRANS}) is tantamount to
 let the current vanish in (\ref{current}), which after integration
  gives system (\ref{deter}) independent of
 $\gamma$, as to be expected.

\section{\bf Conclusion}
The model of the discrete event system presented in this report turned
out very friendly for simulation runs. Although the dimension be small
(curves in the Euclidean plane), several basic phase-transition
phenomena have been observed, among which a glassy phase. As we stove
to point out, there are many ways to describe this system, which bring
to light connections between various stochastic and algebraic formalisms.

Nonetheless, in our opinion, the most efficient way toward concrete
 mathematical and physical properties appears to be on the track of
 coupled exclusion processes. This mapping allows also to address the
 continuous limit considered in the last section, and this approach can
 be a method for coding numerical simulations. In this manner, we have
 been able to observe that when we alternatively freeze one of the
 subsystems, letting the other one reach its equilibrium, the whole process
 attains  its stationary state, and moreover  much quicker.

 With regard to the dynamic,  deeper investigations are needed in
 order to include fluctuations directly into the equations, and
 to  clarify the role of the parameter $\delta$. This would
 give a firm starting point to get an insight into slow dynamics and
 into the non-linear excitations which are swarming in the glassy
 phase (metastable states). Open boundary conditions and presence of
 currents would also be worth investigating. Actually, it could be
 most rewarding (and this not hopeless) to generalize the model to
 higher dimensions and to find related concrete applications, 
 for example in biology (evolution of RNA and proteins).

\medskip
{\bf Acknowledgment}
The authors want to thank  Arnaud de La Fortelle (INRIA)
and  Kirone Mallick (CEA) for many valuable discussions.
The second author  also acknowledges Christophe
Josserand (CNRS) for useful advices,  
especially about numerical experiments.


\begin{thebibliography}{99}
\bibliographystyle{siam}

\bibitem{PGDG} P.G. De Gennes 1979 {\it Scaling Concepts in Polymer
Physics} (Ithaca, NY: Cornell University Press).

\bibitem{EDWARD} M. Doi and S.F. Edwards 1986 {\it The Theory of
Polymer dynamics} (Oxford University Press).
\bibitem{SPITZER} F. Spitzer,  Trans. Am. Math. Soc. {\bf 87} (1958) 187 

\bibitem{LEVY} P. L\'evy 1948 {\it Processus Stochastiques et Mouvement
Brownien} (Gauthier-Villars).

\bibitem{YOR} J.W. Pitman, M. Yor, Ann. Probab. {\bf 17} (1989) 965

\bibitem{BGS} Ch.L. Berthelsen, J.A. Glazier, M.H. Skolnick, Phys.
Rev. A 45 (1992) 8902--8913.

\bibitem{BURGERS} J.M. Burgers, Adv. Appl. Mech. {\bf 1} (1949) 177.

\bibitem{CD} S. Cebrat, M. Dudek {\it The effect of DNA phase
structure on DNA walks}. The Eur. J. Phys. {\bf B} 3 (1998) 271--276.

\bibitem{COMTET} A. Comtet, J. Desbois, S. Ouvry, J. Phys. A: Math.
Gen. {\bf 23} (1990) 5637.

\bibitem{DUR} R. Durrett (1995) {\it Ten lectures on particle systems}.
Lecture Notes in Maths. 1608, Springer, 97--201.

\bibitem{ROUSE} P.E. Rouse,  J. Chem. Phys. {\bf 21}, 1273. (1953)

\bibitem{DESBOIS} O. BÈnichoux, J. Desbois, J. Phys. A. Math. Gen.
{\bf 33} (2000) 6655--6667.

\bibitem{HAKIM} V. Hakim, J.P. Nadal {\it Exact results for 2D
directed animals on a strip of finite width.} J. Phys. A: Math. Gen.
{\bf 16} (1983) L213.

\bibitem{ALMEIDA} J.S. Almeida, J.A. CarriÁo, A. Maretzek, P.A. Noble
and M. Fletcher {\it Analysis of genomic sequences by Chaos Game
Representation} J. of Bioinfo. {\bf 17} 5 (2001) 429--437.

\bibitem{DORFMAN} J.R. Dorfman 1999 {\it An Introduction to Chaos in
Nonequilibrium Statistical Mechanics}. Cambridge University
Press.

\bibitem{KAL} O. Kallenberg (2001) {\it Foundations of Modern
Probability.} Second Edition, Springer.

\bibitem{KPZ} M. Kardar, G. Parisi and Y.C. Zhang, Phys. Rev. Lett.
{\bf 56} (1986) 889.

\bibitem{KOMLOS} J. Komlos, P. Major, G. Tusnady (1975,1976) 
{\it An approximation of partial sums of independent RV's and the
sample DF I,II.} Z. Warsch. verw. Gebiete. {\bf 32}, 111--131.

\bibitem{LIG} T.M. Liggett (1985) {\it Interacting Particle
Systems.} Grundlehren der mathematischen Wissenschaften, Springer.

\bibitem{NOVIKOV} B. Doubrovine, S. Novikov, A. Fomenko (1979) {\it
GÈomÈtrie Contemporaine. 2e partie GÈomÈtrie et Topologie des VariÈtÈs}.
…ditions Mir.

\bibitem{DERRIDA} B. Derrida, M.R. Evans, V. Hakim and V. Pasquier
{\it Exact solution for 1D asymmetric exclusion model using a matrix
formulation} J. Phys. A: Math. Gen. {\bf 26} (1993) 1493--1517.

\bibitem{OKSENDAL} B.K. ÿksendal (1985) {\it Stochastic Differential
Equations}. Springer.



\end{thebibliography}
\end{document}